\documentclass[lettersize, journal]{IEEEtran}
\usepackage{array}
\usepackage{float}
\usepackage{caption}
\usepackage{color}

\usepackage{graphicx} \usepackage{float} %\usepackage{subfigure} 
\usepackage{subcaption}
\usepackage{cite}
\usepackage{amsmath,amssymb,amsfonts}
\usepackage{amsthm}
\usepackage{graphicx}
\usepackage{textcomp}
\usepackage{xcolor}
\usepackage{svg}
\usepackage{algorithm}
\usepackage{algorithmic}
\usepackage{mathtools}
\usepackage{lipsum}

\newtheorem{remark}{Remark}
\newtheorem{theorem}{Theorem}

\newtheorem{lemma}{Lemma}

\newtheorem{corollary}{Corollary}

\newcounter{TempEqCnt}

\allowdisplaybreaks
\makeatletter
\def\ScaleIfNeeded{%
\ifdim\Gin@nat@ width>\linewidth \linewidth \else \Gin@nat@width
\fi } \makeatother

\graphicspath{{Fig/}}
\begin{document}
\title{Near-Field Integrated Sensing and Communications for Secure UAV Networks}
\author{Jingjing Zhao, Songtao Xue, Kaiquan Cai, Xidong Mu, 
Yuanwei Liu,~\IEEEmembership{Fellow,~IEEE}, and Yanbo Zhu
\thanks{J. Zhao, S. Xue, K. Cai, and Y. Zhu are with the School of Electronics and Information Engineering, Beihang University, Beijing, China, and also with the State Key Laboratory of CNS/ATM, Beijing, China. (e-mail:\{jingjingzhao, xuesongtao, caikq, zhuyanbo\}@buaa.edu.cn). }
\thanks{X. Mu is with the Centre for Wireless Innovation (CWI), Queen's University Belfast, Belfast, BT3 9DT, U.K. (e-mail: x.mu@qub.ac.uk).}
\thanks{Y. Liu is with the Department of Electrical and Electronic Engineering, the University of Hong Kong, Hong Kong, China (e-mail: yuanwei@hku.hk).}
}

%\title{Integrated Sensing and Communications for Near-field UAV Secure Communications}
\maketitle

\begin{abstract}
  A novel near-field integrated sensing and communications framework for secure unmanned aerial vehicle (UAV) networks with high time efficiency is proposed. A ground base station (GBS) with large aperture size communicates with one communication UAV (C-UAV) under the existence of one eavesdropping UAV (E-UAV), where the artificial noise (AN) is employed for both jamming and sensing purpose. Given that the E-UAV's motion model is unknown at the GBS, we first propose a near-field  localization and trajectory tracking scheme. Specifically, exploiting the variant Doppler shift observations over the spatial domain in the near field, the E-UAV's three-dimensional (3D) velocities are estimated from echo signals. To provide the timely correction of location prediction errors, the extended Kalman filter (EKF) is adopted to fuse the predicted states and the measured ones. Subsequently, based on the real-time predicated location of the E-UAV, we further propose a joint GBS beamforming and C-UAV trajectory design scheme for maximizing the instantaneous secrecy rate, while guaranteeing the sensing accuracy constraint. To solve the resultant non-convex problem, an alternating optimization approach is developed, where the near-field GBS beamforming and the C-UAV trajectory design subproblems are iteratively solved by exploiting the successive convex approximation method. Finally, our numerical results unveil that: 1) the E-UAV's 3D velocities and location can be accurately estimated in real time with our proposed framework by exploiting the near-field spherical wave propagation; and 2) the proposed framework achieves superior secrecy rate compared to benchmark schemes and closely approaches the performance when the E-UAV trajectory is perfectly known.  

\end{abstract}


\section{Introduction}
With the fast development of the low-altitude economy, unmanned aerial vehicles (UAVs) have received significant attention for potential usage in military, civilian, and commercial applications~\cite{UAV1}. Particularly, benefiting from the high mobility and the low operation cost, UAVs are anticipated to play significant roles in future wireless networks~\cite{UAV2}. In contrast to the terrestrial communications, the ground-to-air communications have higher probability to be dominated by the line-of-sight (LoS) links~\cite{UAVLOS1, site}, which is beneficial for improving the communication quality. Moreover, the UAV's locations can be flexibly adjusted to adapt to the time-varying communications environment, which helps to construct preferable channel conditions in real time. 

Despite the above advantages, one prominent challenge for facilitating UAV communications is the potential eavesdropping risk posed by the inherent broadcast nature of the LoS-dominated channels~\cite{UAVLOS2}.
As such, it is critical to develop physical layer security (PLS) technologies to provide flexible security services for safeguarding UAV communications. Thanks to the capability of the multiple-input multiple-output (MIMO) technique for steering the signals toward the desired directions, the secrecy rate at the receivers can be enhanced via effective beamforming~\cite{MIMOUAV1,MIMOSECURE}. However, since the eavesdroppers are usually non-cooperative nodes, it is impractical to obtain the wiretap channel state information (CSI) with the conventional pilot-based estimation approach~\cite{pilot1}. In addition, due to the high mobility of the UAV, the CSI tracking is a tough task to guarantee stable communication performance while meeting the strict latency requirement. Recently, the concept of sensing-aided communication, which leverages the integrated sensing and communications (ISAC) techniques~\cite{9737357, 9851407}, has emerged for obtaining the communication/wiretap CSI and thereby assisting the subsequent beamforming design~\cite{ISAC1,ISAC2}. This approach eliminates the requirement for transmitting pilot signals, and instead estimating the CSI based on the distance and direction parameters obtained from echo signals reflected by legitimate users/eavesdroppers~\cite{ISAC3}. 

With the increment of the array aperture size and the carrier frequencies, the Rayleigh distance in the next-generation communications system can span up to several tens to hundreds of meters~\cite{NFCMIMO1,NFCMIMO2}. Therefore, the UAVs that move in the low-altitude space are more likely to communicate with the ground base station (GBS) in the near-field region. Unlike the planar wave approximation in the far-field channel model, the more complex spherical wave front brings in new opportunities for the sensing and secure communication design.
%~\cite{NFCWAVE,NFCMIMO3}. 
% On the one hand, the spherical-wave propagation facilitates new degrees of freedom (DoFs) for sensing, through exploiting the variant angular and distance observations over the antennas~\cite{NFCBEAMFOCUS}. 
% On the other hand, since the beams can be focused at a specific location (\textit{beamfocusing}) rather than a specific direction (\textit{beam steering}), this shift introduces higher DoFs for the anti-eavesdropping beamforming design. 
As a result, the sensing-aided secure UAV communications in the near field deserves further investigation.

\subsection{Related Works}
\subsubsection{Secure UAV Communications}
Extensive research works have been devoted to secure UAV communications.
In~\cite{UAVSECURE2}, the UAV navigation problem was investigated for securing the ground-to-air communications in the presence of collaborative eavesdroppers. The authors of~\cite{UAVSECURE1} investigated the average secrecy rate maximization problem via jointly optimizing the robust trajectory and the UAV transmit power, assuming imperfect knowledge of the eavesdropper locations. In~\cite{UAVSECURE3}, the PLS of UAV relaying communication system was studied, where the UAV forwarded the information from the GBS to the ground nodes in the presence of an eavesdropper. The authors of~\cite{UAVJAMMING2} discussed the security and reliability performance of a ground communication system, where a UAV was deployed to send jamming signals to protect against an eavesdropper. 
Considering the multi-UAVs scenario with multiple eavesdroppers, the authors of~\cite{multiuav} leveraged the cooperation among UAVs for simultaneously sending information signals to the legitimate nodes and jamming signals to the eavesdroppers. {In~\cite{9416239}, the reconfigurable intelligent surface (RIS) was applied to enhance the average secrecy rate of the UAV communications under the assumption of the imperfect wiretap CSI. Furthermore, the joint UAV transmit power, trajectory, and RIS beamforming were jointly optimized in~\cite{10288199}, with the aim of maximizing the average secrecy rate in the RIS-aided multi-UAV communications scenario.
The aforementioned works all assumed that the eavesdroppers were static and the wiretap CSI was fully or at least partially known, while the wiretap CSI is not easy to be obtained in practice as the eavesdropper is normally a non-cooperative node. 
\subsubsection{Sensing-Aided Communications}
% The radar and pilot transmission may cause overhead, particularly in high-mobility scenarios. 
% In this context, the integrated sensing and communication (ISAC)
% techniques have emerged as a promising solution~\cite{ISAC1,ISAC2,ISAC3}. By exploiting
% the echo signals for sensing, the position parameters can
% be estimated to reconstruct CSI, where pilot-based channel estimation is no longer required. This approach not only enhances spectral efficiency but also facilitates the real-time acquisition of dynamic CSI, thereby ensuring more effective resource management and better adaptation to rapidly changing environments.  
To reap the benefits of the ISAC technique, the sensing-aided communications have attracted some research contributions recently, especially in the far-field scenario. For instance, the authors of~\cite{ISACFAR1} proposed to jointly design the transmit beamformers of communication and radar systems to minimize the maximum eavesdropping signal-to-interference-plus-noise ratio (SINR). In~\cite{ISACFAR2}, the authors proposed a beamforming optimization method for downlink secure ISAC systems, where the communication secrecy and target sensing performance were well balanced under imperfect eavesdropper CSI.  To further enhance the communications security, the authors of~\cite{ISACFAR3} employed a combined Capon and approximate maximum likelihood (CAML) technique to search for the potential eavesdroppers' directions and formulated a weighted optimization problem to enhance the secrecy rate while minimizing the eavesdroppers’ Cramér-Rao Bound (CRB). In addition, the sensing-assisted predictive beamforming has been investigated for high-dynamic scenarios~\cite{ISACli2,ISACli1}. The authors of~\cite{ISACli2} proposed a sensing-aided beamforming prediction algorithm based on the extended Kalman filter (EKF), where the ISAC echo signals reflected by vehicles were employed to predict the CSI for time-varying channels. {Furthermore, the authors of~\cite{ISACli1} 
developed a two-stage beamforming scheme, where the CSI was estimated from echo signals and predicted via a neural network, which was followed by the joint active and passive beamforming. In~\cite{weizhiqiang}, an ISAC framework was proposed for safeguarding the UAV communications against a mobile eavesdropper, where the artificial noise (AN) was used for estimating the wiretap channel so as to assist the online UAV navigation and resource allocation design. }

Recently, some works have also started to explore the near-field target sensing to assist the reliable communications. {For example, the authors of~\cite{NFC-LOCA2} investigated the maximum likelihood (ML) approach to enhance the near-field localization performance in the presence of noise.} To ensure reliable communication performance, the authors of~\cite{10288339} investigated the beam training and tracking schemes for the extremely large-scale MIMO systems with partially-connected combiner structures, where the kinematic model was adopted to characterize the near-field channel variations.} In~\cite{senseaid}, the authors proposed a near-field beamforming optimization algorithm that utilized environmental sensing information to minimize BS transmit power while realizing precise multi-target detection.
Considering high-dynamic scenarios on the ground, the authors of~\cite{NFC_V} proposed a motion tracking method that leveraged the characteristics of spherical waves to sense the two-dimensional (2D) velocity of the target, for assisting the predictive beamforming. 

\vspace{-0.3cm}
\subsection{Motivations and Contributions}
{The existing works on sensing-aided UAV secure communications only focus on the far-field region, while there is a paucity of investigations on the near field. Compared to the far field, the shift on near-field propagation characteristics introduces new opportunities for the time-sensitive secure UAV communications design as follows.
\begin{itemize}
    \item For sensing, the variant Doppler observations over different antennas opens up the possibility for simultaneously estimating the three-dimensional (3D) velocities of the mobile eavesdropper, which facilitates the eavesdropper location prediction without the prior knowledge of the motion model. This is of vital importance for UAV secure communications to localize high-mobility eavesdropping UAVs.
    \item For communications, near-field propagation enables the unique \emph{beamfocusing} capability, i.e., targeting the signal on a specific location in the near field instead of a specific direction in the far field. This provides enhanced flexibility to prevent legitimate information leakage and thus improve anti-eavesdropping performance, which is beneficial for secure UAV communications.
\end{itemize}}

Despite the above attractive benefits, the spherical wave-based near-field propagation causes the sensing and communications design quite complicated. The sensing and anti-eavesdropping beamforming schemes have to be re-designed to unlock the great potential of near field for secure UAV communications. To the best of the authors' knowledge, the design of near-field secure UAV communications with the aid of 3D sensing technologies has not been well studied yet, which motivates us to contribute this work.

Against the above background, this paper investigates the near-field sensing-assisted secure UAV communications framework, to guarantee PLS for the communication link from the ground base station (GBS) to a communication UAV (C-UAV) in the presence of an eavesdropping UAV (E-UAV).  Benefiting from the ISAC technique, the C-UAV trajectory and the GBS beamforming are designed online with low latency based on the sensing results obtained with the echo signals. 
For the sake of brevity, the main contributions of this paper are summarized as follows.

\begin{itemize}
\item We consider the secure UAV communication scenario, where the GBS transmits information signals toward the C-UAV and also sends artificial noise (AN) signal toward the E-UAV for both sensing and jamming purpose. To realize the online PLS design without the prior knowledge of the E-UAVs's motion model, we propose a sensing-aided secure communication framework, which consists of two schemes, namely the \textit{E-UAV trajectory tracking scheme} and the \textit{joint GBS beamforming and C-UAV trajectory design scheme}. 

\item For the E-UAV trajectory tracking, we first propose a ML method to estimate the E-UAV's 3D velocities by exploiting the spatial variation of Doppler shifts in the near field. Furthermore, we employ the EKF-based method to fuse the location predicted with the sensed 3D velocities and that estimated with the echoes, for improving the E-UAV trajectory tracking accuracy. 

\item For the joint GBS beamforming and C-UAV trajectory design, we formulate an instantaneous secrecy rate maximization problem with the constraint on the GBS transmit power and the E-UAV tracking accuracy. To solve the resultant non-convex problem, an alternating optimization (AO) algorithm is developed to decouple the optimization variables, where the GBS beamforming and the C-UAV trajectory design subproblems are alternately solved by employing the successive convex approximation (SCA) method. 

\item Our numerical results unveil that 1) the proposed E-UAV trajectory tracking scheme achieves high-accuracy localization of the E-UAV's flight route in real time; and 2) the achievable secrecy rate of the proposed sensing-aided secure UAV communications framework closely approaches to that of the case where the E-UAV trajectory is perfectly known.
\end{itemize}
\subsection{Organization and Notations} 
The remainder of this paper is organized as follows.
Section II introduces the near-field sensing-aided secure UAV communication system model. Section III
proposes the E-UAV 3D trajectory tracking scheme, which is followed by the AO-based joint GBS beamforming and C-UAV trajectory design in Section IV. Section V presents the simulation results, and Section V concludes the paper.

\textit{Notations:}
Vectors and matrices are denoted by boldface lowercase and boldface capital letters, respectively. $\mathbb{C}^{M\times N}$ denotes the space of all $M\times N$
matrices with complex entries. $\mathbf{A}^*$, $\mathbf{A}^T$ and $\mathbf{A}^H$ denote the conjugate,  transpose, and conjugate transpose of the matrix $\mathbf{A}$, respectively. $\mathbf{A}_{i,j}$ returns the entry in the $i$-th row and $j$-th column of matrix $\mathbf{A}$. $\mathrm{tr}(\mathbf{A})$
and $\mathrm{rank}(\mathbf{A})$ denote the trace and the rank of matrix $\mathbf{A}$. $\odot$ is the Hadamard product. $\|\cdot\|$ denotes the norm of the vector. $\|\cdot\|_F$ denotes the Frobenius norm of a matrix. $\mathcal{CN}\left(u,v\right)$  denotes a Gaussian distribution 
with mean $m$ and variance $v$. $|\cdot|$ denotes the absolute value of a complex scalar. $\mathbf{I}_M$
denotes the $M \times M$ identity matrix. $\mathbf{0}_M$
denotes the $M \times M$ zero matrix. $[x]^{+}$ stands for
$\max\{x, 0\}$.

\section{System Model and Problem Formulation}
%\vspace{-0.3cm}
%\subsection{Signal Model}
%\label{sec:active STAR-RIS}
%We establish the signal model of the active STAR-RIS. Assume that $x_{\text{R},m}$ is the incident signal onto the $m$-th element of the 

% \vspace{-0.3cm}
%\vspace{-0.5cm}

\begin{figure}
    \centering    \includegraphics[width=1\linewidth]{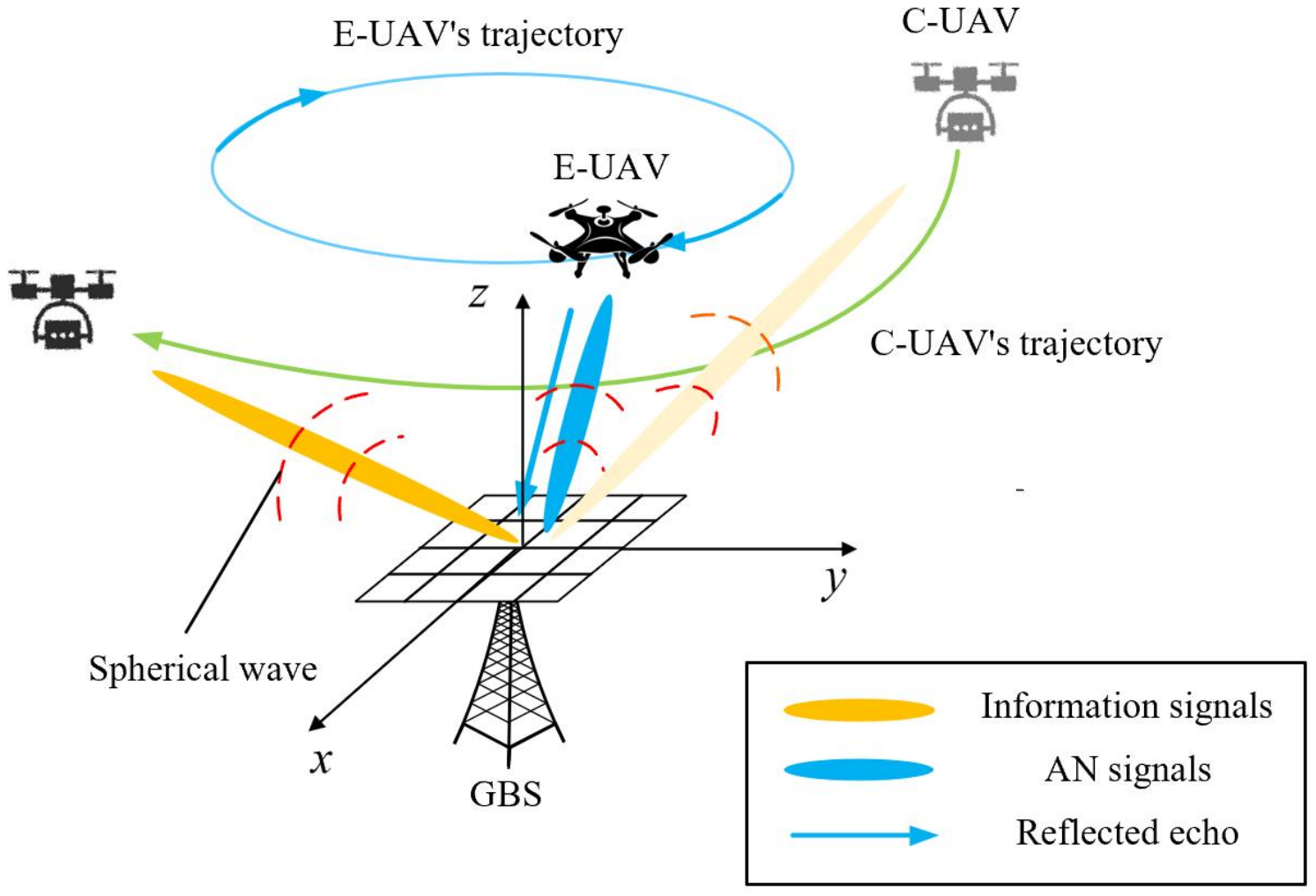}
    \caption{Illustration of the near-field UAV communication scenario with the presence of an eavesdropper.}
    \label{fig:sys}
\end{figure}

{As shown in Fig.~\ref{fig:sys}, we consider a near-field\footnote{{We assume that both the C-UAV and E-UAV are in the near-field region during the whole flight.}} UAV communication system that consists of one ISAC GBS, one C-UAV receiving signals from the GBS, and one E-UAV wiretapping the GBS information.} {Thanks to few scatterers in the air-to-ground communication environment, we assume the LoS-dominant case for channels from the GBS to the C-UAV and the E-UAV~\cite{UAVSECURE1,10288199}.} The GBS is deployed with a rectangular uniform planar array (UPA) that consists of $M=M_x\times M_y$ antennas with the spacing $d$, while the C-UAV and the E-UAV are both equipped with a single antenna. The index of the $m$-th GBS antenna is given by $m=aM_y+b$, where $a=0,\dots, M_x-1$ and $b=1,\dots, M_y$. We assume that the GBS operates in the full-duplex mode, which allows it to transmit and receive signals simultaneously. With the aid of advanced self-interference mitigation
 techniques~\cite{DULL}, we assume that the full-duplex GBS can realize perfect self-interference mitigation.
The bandwidth of the system, denoted by $B$, is related to the symbol duration $T_s=\frac{1}{B}$.
Let $N$ denote the number of symbol durations in a single coherent processing interval (CPI) with the duration of $\Delta t$ for the near-field sensing, during which the UAVs' location and velocity are assumed to be unchanged. We set a Cartesian coordinate system for which the origin is chosen as the center of the GBS UPA and the $x$ and $y$ axes are aligned with the column and row of the UPA, respectively, while the $z$ axis points up to complete a right handed coordinate frame. As such, the coordinate of the UPA center is given by $\mathbf{q}_b = [0,0,0]^T$, while that of the $m\mathrm{\text{-}th}$ antenna is denoted by $\mathbf{q}_{m}=[x_{m},y_{m}, 0]^T$.  In the $l\in\{1,..., L\}\mathrm{\text{-}th}$ CPI, the position and the velocity of the E-UAV are denoted by $\mathbf{q}_ {e,l}=\left[x^e_{l},y^e_{l},z^e_{l}\right]^T$ and $\mathbf{v}_ {e,l}=[{v}^x_{e,l},{v}^y_{e,l},{v}^z_{e,l}]^{T}$, respectively, and the position of the C-UAV is represented by $\mathbf{q}_{c,l}=\left[x^{c}_{l},y^{c}_{l},z^{c}_{l}\right]^T$. Given that the E-UAV trajectory is unknown at the GBS, the GBS sends artificial noise (AN) signals to the E-UAV for both sensing and jamming purpose. Then, the transmitted signal $\mathbf{x}_{l}(n)$ of the GBS in the time slot $n$ of the $l$-th CPI is given by
\begin{equation}
\label{eq: transmit signal}
\mathbf{x}_{l}(n)=\mathbf{w}^{c}_{l}s_{l}(n)+\mathbf{w}^e_{l}a_{l}(n),
\end{equation}
where $s_{l}(n)\sim\mathcal{CN}\left(0,1\right)$ and $a_{l}(n)\sim\mathcal{CN}\left(0,1\right)$ represent the information and AN signals, respectively, and $\mathbf{w}^{c}_{l}\in\mathbb{C}^{M\times1}$ and $\mathbf{w}^e_{l}\in\mathbb{C}^{M\times1}$ denote the precoding vectors for the information and AN signals in the $l$-th CPI, respectively. Note that, considering the tradeoff between the system performance and the computational complexity, the same beamformers are utilized within each CPI. 

% \textcolor{red}{Denote the GBS transmitted signal by $\mathbf{x}(n)=[x_{1}(n),\cdots,x_{m}(n)]\in\mathbb{C}^{M\times1}$, where $x_{m}(n)$ denotes the transmitted signal on the $m$-th antenna.}
\setcounter{TempEqCnt}{\value{equation}} % 保存当前公式序号到TempEqCnt
\setcounter{equation}{3} % 手动设置当前公式编号为3
\begin{figure*}[ht]
    \begin{equation}
          \label{eq:r_m}
        \bar{r}^e_{l,m}=\sqrt{
    \begin{array}{l}
                       {\left(r^{e}_{l}\right)}^2+\left(a-\frac{M_x-1}{2}\right)^2d^2+\left(b-\frac{M_y+1}{2}\right)^2d^2
                      -2r^{e}_{l}\left(a-\frac{M_x-1}{2}\right)dS_{\theta^{e}_{l}}C_{\phi^{e}_{l}}-2r^{e}_{l}\left(b-\frac{M_y+1}{2}\right)dC_{\theta^{e}_{l}}C_{\phi^{e}_{l}}
              \end{array}       
           }.
    \end{equation}
    \hrulefill
\end{figure*}
\addtocounter{equation}{1}

\subsection{Near-Field Sensing Signal Model}
\label{sec:sensing-model}
In practice, the signal transmitted by the GBS is partially received by the E-UAV and partially reflected back to the GBS.  Note that the signals reflected by the C-UAV and other scatterers in the environment are assumed to be effectively suppressed at the GBS by exploiting existing clutter suppression techniques~\cite{zaboyizhi}. In the $n$-th time slot of the $l$-th CPI, the echo signal received by the $m$-th GBS antenna is given by
% To explicitly demonstrate the various observations over each antenna, we express the $m$-th element of $\mathbf{y}^b_{l}(n)$ as
\setcounter{equation}{1}
 \begin{align}
\label{eq:echo}
y^b_{l,m}(n)=&\sum_{i=1}^{M}\frac{\beta}{\left(r_l^e\right)^2}e^{-j\frac{2\pi}{\lambda}\left({r}^e_{l,m}\left(nT_{s}\right)+{r}^e_{l,i}\left(nT_{s}\right)\right)}x_{l,i}(n)\notag
\\&+z^b_{l,m}(n),
 \end{align}
where $r_{l,m}^e(t)=\bar{r}_{l,m}^e+v_{l,m}^et$ represents the time-variant distance from the $m$-th antenna to the E-UAV, $\bar{r}_{l,m}^e$ denotes the distance from the $m$-th antenna to the E-UAV at the beginning of the $l$-th CPI, $v_{l,m}^e$ denotes the E-UAV's  radial velocity component w.r.t. the $m\text{-}\mathrm{th}$ antenna in the $l$-th CPI. Note here that, although we assume the E-UAV's location is constant within one CPI for large-scale fading parameters discussions, such as the channel gain, we take the time-variant delay into consideration in one CPI for facilitating the Doppler estimation as discussed later. Still referring to~\eqref{eq:echo}, $\beta$ is a constant factor given by $\beta =\beta_0\sqrt{\frac{\sigma_\mathrm{RCS}}{4\pi}}$, 
with $\beta_0$ and $\sigma_{\mathrm{RCS}}$ representing the channel gain at the reference distance of $1$~m and the radar cross-section of the E-UAV, respectively.
Furthermore, {${x}_{l,i}(n)={w}^c_{l,i}s_{l}(n)+{w}_{l,i}^ea_{l}(n)$ denotes the transmitted signal of the $i$-th antenna, ${w}^c_{l,i}$ and ${w}_{l,i}^e$ represent the $i$-th element of $\mathbf{w}_{l}^c$ and $\mathbf{w}_{l}^e$, respectively},
and $z_{l,m}^b(n)\sim\mathcal{CN}\left(0,\sigma_b^2\right)$ denotes the additive white Gaussian noise (AWGN) at the $n$-th antenna. 

Denote the array response vector at the GBS for the E-UAV at the beginning of the $l$-th CPI as $\boldsymbol{\alpha}(r^{e}_{l}, \theta^{e}_{l}, \phi^{e}_{l})$, where $r^{e}_{l}$, $\theta^{e}_{l}$ and $\phi^{e}_{l}$ represent the distance, azimuthal angle, and elevation angle between the E-UAV and the GBS UPA center in the $l\text{-}\mathrm{th}$ CPI, respectively. Specifically, $\boldsymbol{\alpha}(r^{e}_{l}, \theta^{e}_{l}, \phi^{e}_{l})$ is given by
\begin{equation}
\label{eq: Tx steering vector}
\begin{aligned}
\boldsymbol{\alpha}(r^{e}_{l}, &\theta^{e}_{l}, \phi^{e}_{l})=\left[{e^{-j{\frac{2\pi}{\lambda}}\bar{r}^e_{l,1}}},...,{e^{-j{\frac{2\pi}{\lambda}}\bar{r}^e_{l,M}}}\right]^T.
\end{aligned}
\end{equation}
Based on the geometric relationship, the expression of $\bar{r}^e_{l,m}$ with respect to (w.r.t.) $r^{e}_{l}, \theta^{e}_{l}$ and $\phi^{e}_{l}$ is given in $\eqref{eq:r_m}$, where $S_{\theta^e_l}=\sin{\theta^e_l}$, $C_{\theta^e_l}=\cos{\theta^e_l}$ and $C_{\phi^e_l}=\cos{\phi^e_l}$. Further, denote the Doppler frequency shift over the $m$-th antenna in the $l$-th CPI by $f^D_{l,m}$. Then, we define the time-variant phase-Doppler vector as $\mathbf{d}_{l,n}\left(\mathbf{f}^{D}_l\right)$, with $\mathbf{f}_l^{D}=\left[f_{l,1}^D, ..., f_{l,M}^D\right]$, in the $n$-th time slot of the $l$-th CPI as
\setcounter{equation}{4}
\begin{equation}
\label{eq:phase-Doppler}
\mathbf{d}_{l,n}\left(\mathbf{f}_l^D\right)=\left[e^{-j2\pi f_{l,1}^DnT_s},\ldots, e^{-j2\pi f_{l,M}^DnT_s}\right]^T.
\end{equation}

{Based on~\eqref{eq:echo}-\eqref{eq:phase-Doppler}, the received echo signals over all the GBS antennas in the $n$-th time slot of the $l$-th CPI, i.e., $\mathbf{y}^{b}_{l}(n) = \left[y^b_{l,1}\left(n\right),\ldots,y^b_{l,M}\left(n\right)\right]^T\in \mathbb{C}^{M\times 1}$, can be expressed as 
\begin{align}
 \label{eq: all echo}
\mathbf{y}^{b}_{l}(n)=\frac{\beta}{\left(r_l^e\right)^2}\mathbf{A}\left(r^{e}_{l},\theta^{e}_{l},\phi^{e}_{l}\right)\odot\mathbf{D}_{l,n}\left(\mathbf{f}^D_{l}\right)\mathbf{x}_{l}(n)+\mathbf{z}_l^b(n),
\end{align}
where $\mathbf{A}\left(r^{e}_{l},\theta^{e}_{l},\phi^{e}_{l}\right)=\boldsymbol{\alpha} \left(r^{e}_{l}, \theta^{e}_{l} , \phi^{e}_{l} \right)\times\boldsymbol{\alpha}^T\left(r^{e}_{l} , \theta^{e}_{l} , \phi^{e}_{l}\right)$, $\mathbf{D}_{l,n}\left(\mathbf{f}_l^D\right)=\mathbf{d}_{l,n}\left(\mathbf{f}_l^D\right)\times\mathbf{d}_{l,n}^T\left(\mathbf{f}_l^D\right)$, {$\mathbf{x}_l(n)=[x_{l,1}(n),\cdots,x_{l,m}(n)]^T\in\mathbb{C}^{M\times1}$, and $\mathbf{z}_l^b(n)=[z_{l,1}^b(n),\cdots,z_{l,m}^b(n)]^T\in\mathbb{C}^{M\times1}$}, respectively.

\begin{remark}
    Different from the far-field scenario, due to the different azimuth and elevation angles observed at each GBS antenna, the Doppler frequency shift varies in the spatial domain for the near-field propagation. 
\end{remark}

\subsection{Near-Field Communication Signal Model}
\label{sec:communication model}
The received signal at the C-UAV in time slot $n$ of the $l$-th CPI is given by
\begin{equation}
\label{eq: receive communication signal}
y^{c}_{l}(n)=\left(\mathbf{h}^{bc}_{l}\right)^H\mathbf{x}_l(n)+z^{c}_{l}(n),
\end{equation}
where $\mathbf{h}^{bc}_{l}$ and $z^{c}_{l}(n)\sim\mathcal{CN}\left(0,\sigma_c^{2}\right)$ denote the channel between the GBS and the C-UAV, and the AWGN at the C-UAV, respectively. Specifically, $\mathbf{h}^{bc}_{l}$ is expressed as
\begin{equation}
\mathbf{h}^{bc}_{l}=\frac{\beta_0}{r_l^c}\boldsymbol{\alpha}(r^{c}_{l},\theta^{c}_{l},\phi^{c}_{l}),
\end{equation}
where $r^{c}_{l}, \theta^{c}_{l}, \phi^{c}_{l}$ represent the distance, azimuthal angle, and elevation angle between the C-UAV and the GBS UPA center in the $l\text{-}\mathrm{th}$ CPI, respectively. {We assume that the C-UAV CSI, i.e., $\mathbf{h}^{bc}_{l}$, is known at the GBS by employing the near-field channel estimation methods~\cite{10078317}.} Accordingly, the achievable data rate at the C-UAV in the $l$-th CPI is given by 
\begin{equation}
\label{eq: communication rate C-UAV}
R^{c}_{l}=\log_2\biggl(1+\frac{\left|\left(\mathbf{h}^{bc}_{l}\right)^H\mathbf{w}^{c}_{l}\right|^2}{\left|\left(\mathbf{h}^{bc}_{l}\right)^H\mathbf{w}^{e}_{l}\right|^2+\sigma_c^2}\biggr).
\end{equation}
Similarly, the leakage information rate at the E-UAV in the $l$-th CPI is given by
\begin{equation}
\label{eq: leakage information rate}
R^{e}_{l}=\log_{2}\left(1+\frac{\left|\left(\mathbf{h}^{be}_{l}\right)^H\mathbf{w}^{c}_{l}\right|^{2}}{\left|\left(\mathbf{h}^{be}_{l}\right)^H\mathbf{w}^{e}_{l}\right|^{2}+\sigma_{e}^{2}}\right),
\end{equation}
where $\mathbf{h}^{be}_{l}$ and $\sigma_{e}^{2}$ denote the channel between the GBS and the E-UAV, and the AWGN power at the E-UAV, respectively. Specifically, $\mathbf{h}^{be}_{l}$ is expressed as 
\begin{equation}
\label{eq:GBS-E-channel}
\mathbf{h}^{be}_{l}=\frac{\beta_0}{r_{l}^{e}}\boldsymbol{\alpha}(r^{e}_{l}, \theta^{e}_{l}, \phi^{e}_{l}).
\end{equation}
Based on $\eqref{eq: communication rate C-UAV}$ and $\eqref{eq: leakage information rate}$, the secrecy rate in the $l$-th CPI can be expressed as follows:
\begin{equation}
\label{eq:secure rate C-UAV}
R^s_{l}=\left[R^{c}_{l}-R^{e}_{l}\right]^{+}.
\end{equation}

The secrecy rate given in~\eqref{eq:secure rate C-UAV} highly depends on the real-time channel (location) of the E-UAV, as shown in~\eqref{eq: Tx steering vector}. However, the E-UAV's location is generally unknown at the GBS as it is a non-cooperative node. As such, we first propose an E-UAV trajectory tracking scheme in Section III. Then, with the aim of maximizing the secrecy rate, we advocate a joint GBS beamforming and C-UAV trajectory design scheme in Section IV, which is based on the estimated E-UAV's real-time locations in Section III. 

% \vspace{-0.2cm}
\section{E-UAV 3D Trajectory Tracking Scheme}
\label{3}
In this section, we propose a near-field E-UAV's 3D velocities sensing and trajectory tracking scheme. Specifically, a ML-based 3D velocities sensing approach is first proposed, which exploits the variant Doppler shift observations over different antennas. Then, an EKF-based state tracking method is proposed to fuse the predicted states and the estimated ones of the E-UAV for improving the tracking accuracy. 

\subsection{3D Velocities Sensing and Localization}

Based on the echo signal model in $\eqref{eq: all echo}$, 
the unknown parameters $\left\{r^e_l, \theta^e_l, \phi^e_l\right\}$ can be estimated by the matched-filtering principle as follows~\cite{MF}: 
\begin{align}
\label{eq:MF}
   &\left\{\hat{r}^e_l,\hat{\theta}^e_l,\hat{\phi}^e_l\right\}\notag\\&=\arg\max_{r^e_l,\theta^e_l,\phi^e_l}\frac{1}{N}\sum_{n=1}^N\boldsymbol{\alpha}^H\left(r^e_l,\theta^e_l,\phi^e_l\right)\mathbf{y}^{b}_l(n)a^{*}_l(n),
   \end{align}
where $\hat{r}^e_l,\hat{\theta}^e_l$, and $\hat{\phi}^e_l$ denote the estimated distance, azimuth angle, and elevation angle from the GBS to the E-UAV, respectively. Note that we assume the AN and the information signals are uncorrelated i.e., $\frac{1}{N}\sum_{n=1}^Ns_l(n)a_l^*(n)\approx0$.

To capture the complete motion of the E-UAV, the real-time velocity sensing is essential. Aggregating $\mathbf{y}^b_l\left(n\right), n\in\left\{1, ..., N\right\}$ into the matrix form as $\mathbf{Y}^{b}_l=\begin{bmatrix}\mathbf{y}_l^b\left(1\right),\mathbf{y}_l^b\left(2\right),\ldots,\mathbf{y}_l^b\left(N\right)\end{bmatrix}$, we can obtain
\begin{equation}
\label{eq:MLE Y}
\mathbf{Y}^{b}_l=\beta\mathbf{X}_l\left(\mathbf{f}^D_l \right)+\mathbf{Z}_l^b,
\end{equation}
% \vspace{-0.4cm}
where $\mathbf{X}_l\left(\mathbf{f}^D_l \right)=\left[\mathbf{G}_l(1)\mathbf{x}_l(1),\dots,\mathbf{G}_l(N)\mathbf{x}_l(N)\right]$, $\mathbf{G}_l(n)=\mathbf{A}(r^e_l,\theta^e_l,\phi^e_l)\odot\mathbf{D}_{l,n}(\mathbf{f}_l^D)$, and
$\mathbf{Z}_l^b=\left[\mathbf{z}_l(1),\mathbf{z}_l(2),\ldots,\mathbf{z}_l(N)\right]$. {As shown in Fig.~\ref{3D_velocity}, to facilitate the 3D velocities derivation, we denote the E-UAV's velocity in the spherical coordinate system that is centered at the GBS UPA center, as $\tilde{\mathbf{v}}_{e,l}=\left[v_{e,l}^{r},v_{e,l}^{\theta}, v_{e,l}^{\phi}\right]$, where $v_{e,l}^{r}$, $v_{e,l}^{\theta}$, and $v_{e,l}^{\phi}$ represent the radial velocity, the azimuthal velocity, and the elevation velocity, respectively. Notably, compared to the 2D velocity sensing, the 3D velocity sensing involves more variables, and thus leads to higher complexity.} {As shown in Fig.~\ref{projection_v}, the mathematical relationship between $\mathbf{f}_l^D$ and $\tilde{\mathbf{v}}_{e,l}$ is expressed as
\begin{equation}
\label{eq:doppler-calculation}
    f_{l,m}^D = \frac{1}{\lambda}\left(v^r_{e,l,m}+v^{\theta}_{e,l,m}+v^{\phi}_{e,l,m}\right).
\end{equation}
{In Eq.~\eqref{eq:doppler-calculation}, $v^r_{e,l,m}$, $v^{\theta}_{e,l,m}$, and $v^{\phi}_{e,l,m}$ denote the projection of the E-UAV's radial velocity, azimuthal velocity, and elevation velocity along the line connecting the E-UAV and the $m$-th GBS antenna, respectively.}} {Thus, we have 
\begin{subequations}
\label{eq:v-3D}
\begin{equation}
\label{eq:vr}
v^r_{e,l,m}=\cos{\alpha_{r,m}}v^r_{e,l},
\end{equation}
\begin{equation}
\label{eq:vt}
v^{\theta}_{e,l,m}=\cos\alpha_{\theta,m}v^{\theta}_{e,l},
\end{equation}
\begin{equation}
\label{eq:vP}
v^{\phi}_{e,l,m}=\cos{\alpha_{\phi,m}}v^{\phi}_{e,l},
\end{equation}
\end{subequations}}

\noindent{where $\alpha_{r,m}$, $\alpha_{\theta,m}$ and $\alpha_{\phi, m}$ represent the included angles between the line connecting the $m$-th GBS antenna and the E-UAV, and the E-UAV's radial, azimuthal, and elevation velocities, respectively.} These angles are illustrated in  Fig.~\ref{projection_v}.
\begin{figure}[htbp]
	\centering
	\begin{subfigure}{0.42\linewidth}
		\centering
		\includegraphics[width=1\linewidth]{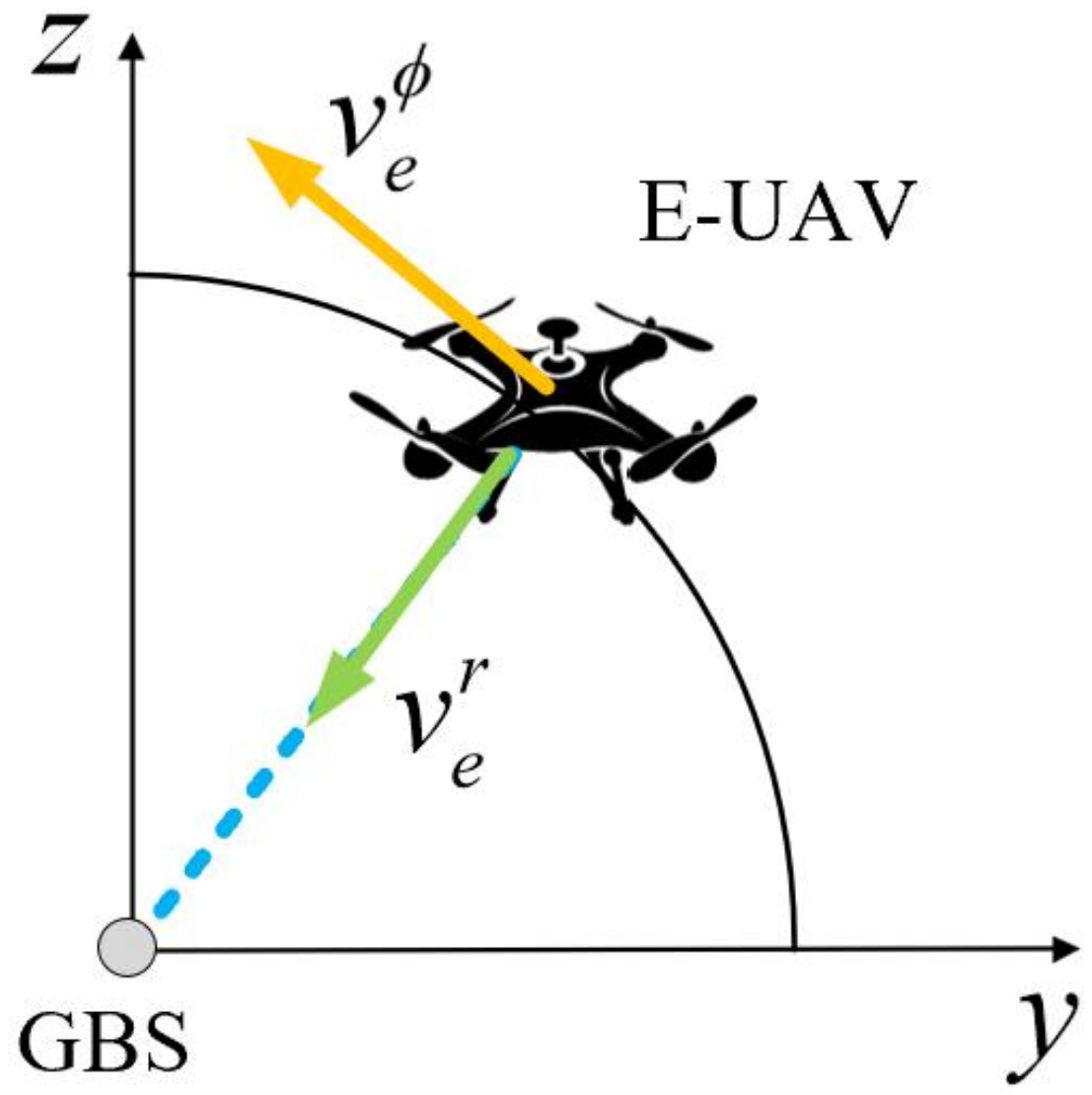}
		\caption{The E-UAV's radial and elevation velocities.}
		\label{chutian3}%文中引用该图片代号
	\end{subfigure}
	\centering
	\begin{subfigure}{0.48\linewidth}
		\centering
		\includegraphics[width=1\linewidth]{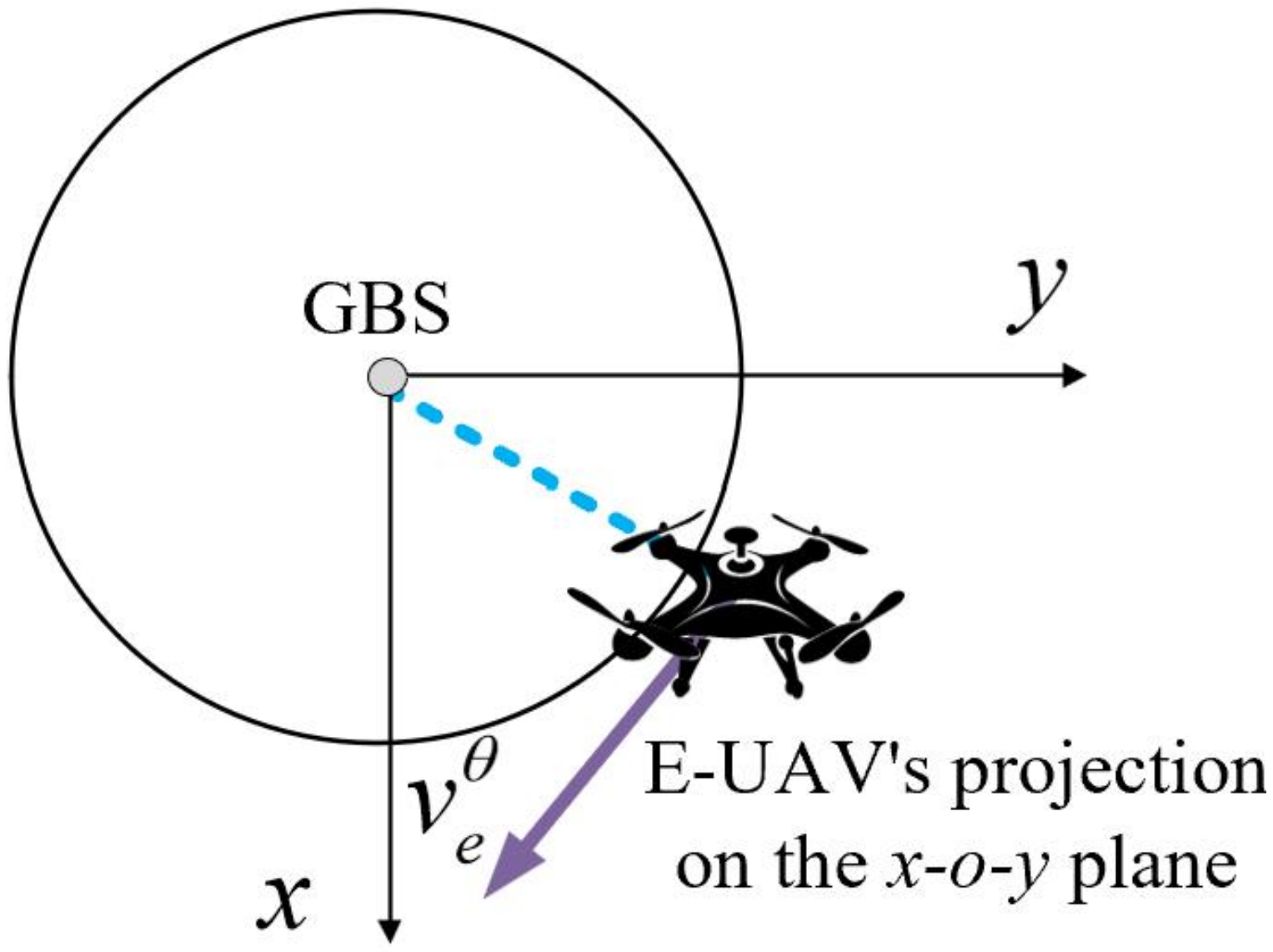}
		\caption{The E-UAV's azimuthal velocity.}
		\label{chutian3}%文中引用该图片代号
	\end{subfigure}
    \caption{Illustration of the E-UAV 's 3D velocities.}
	\label{3D_velocity}
\end{figure}
\begin{figure*}[htbp]
	\centering
	\begin{subfigure}{0.22\linewidth}
		\centering	\includegraphics[width=1\linewidth]{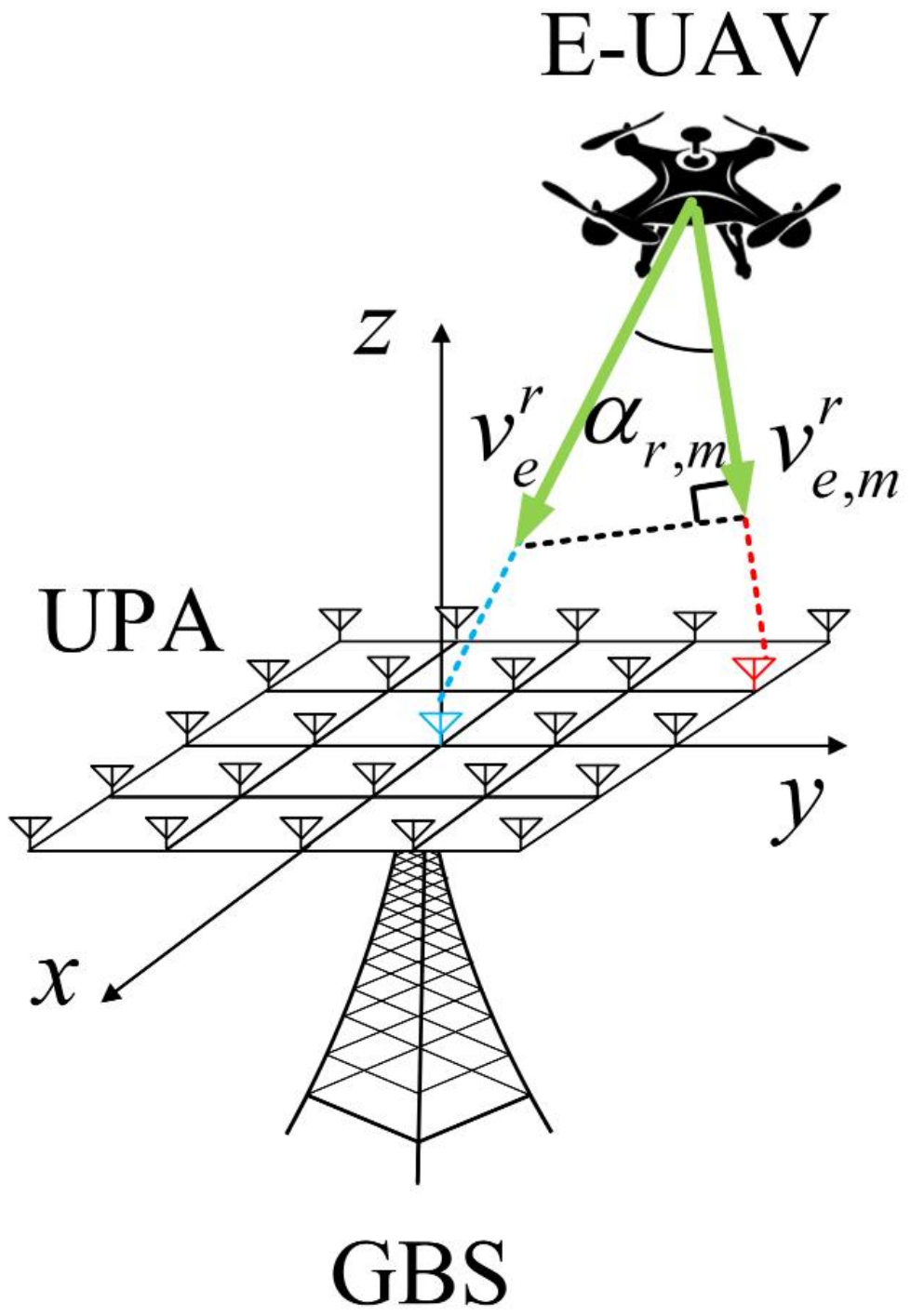}
		\caption{}
		\label{chutian3}%文中引用该图片代号
	\end{subfigure}
     \hspace{12mm}
	\centering
	\begin{subfigure}{0.22\linewidth}
		\centering
		\includegraphics[width=1\linewidth]{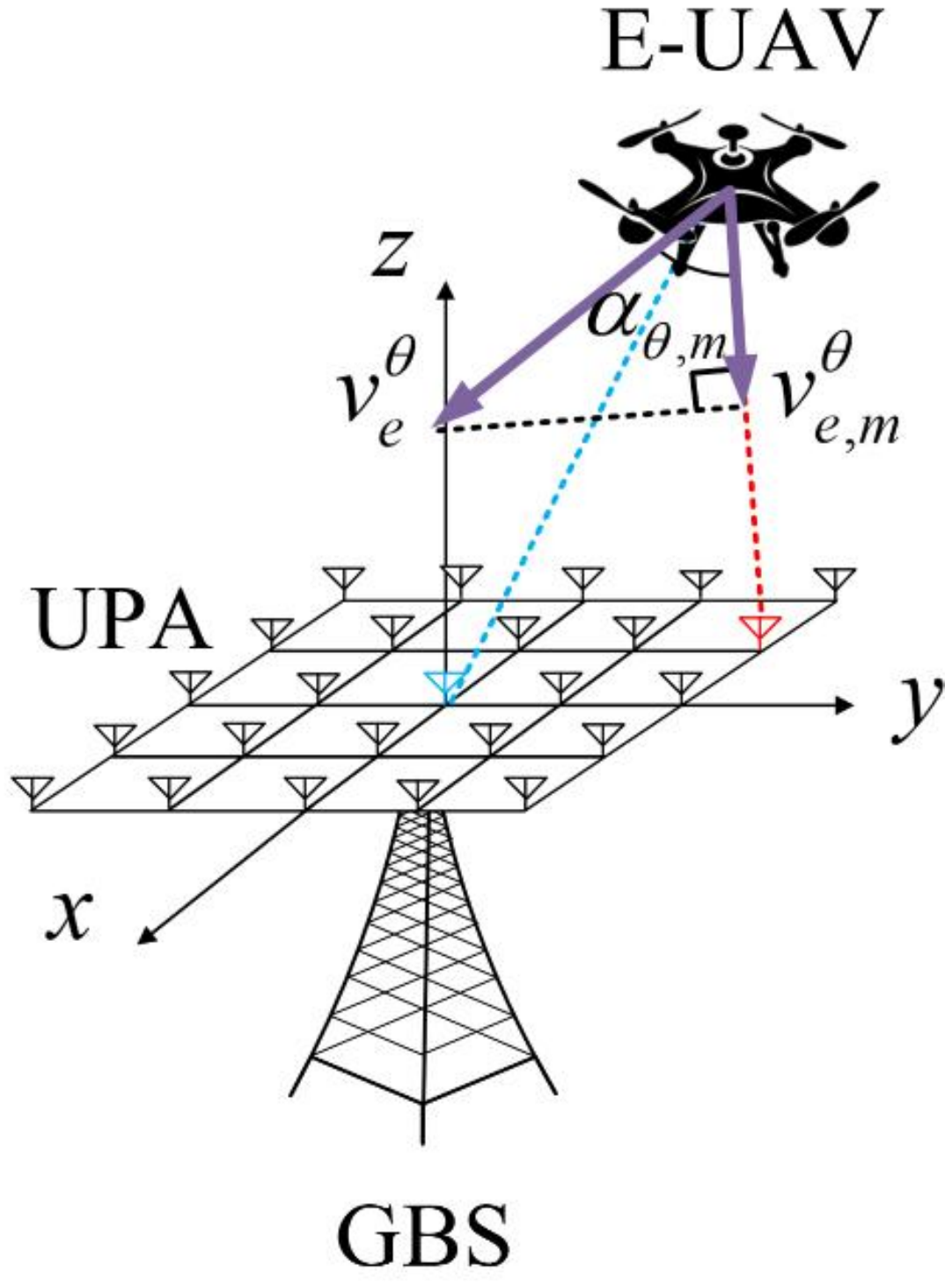}
		\caption{}
		\label{chutian3}%文中引用该图片代号
	\end{subfigure}
    \hspace{12mm}
	\centering
	\begin{subfigure}{0.22\linewidth}
		\centering
		\includegraphics[width=1\linewidth]{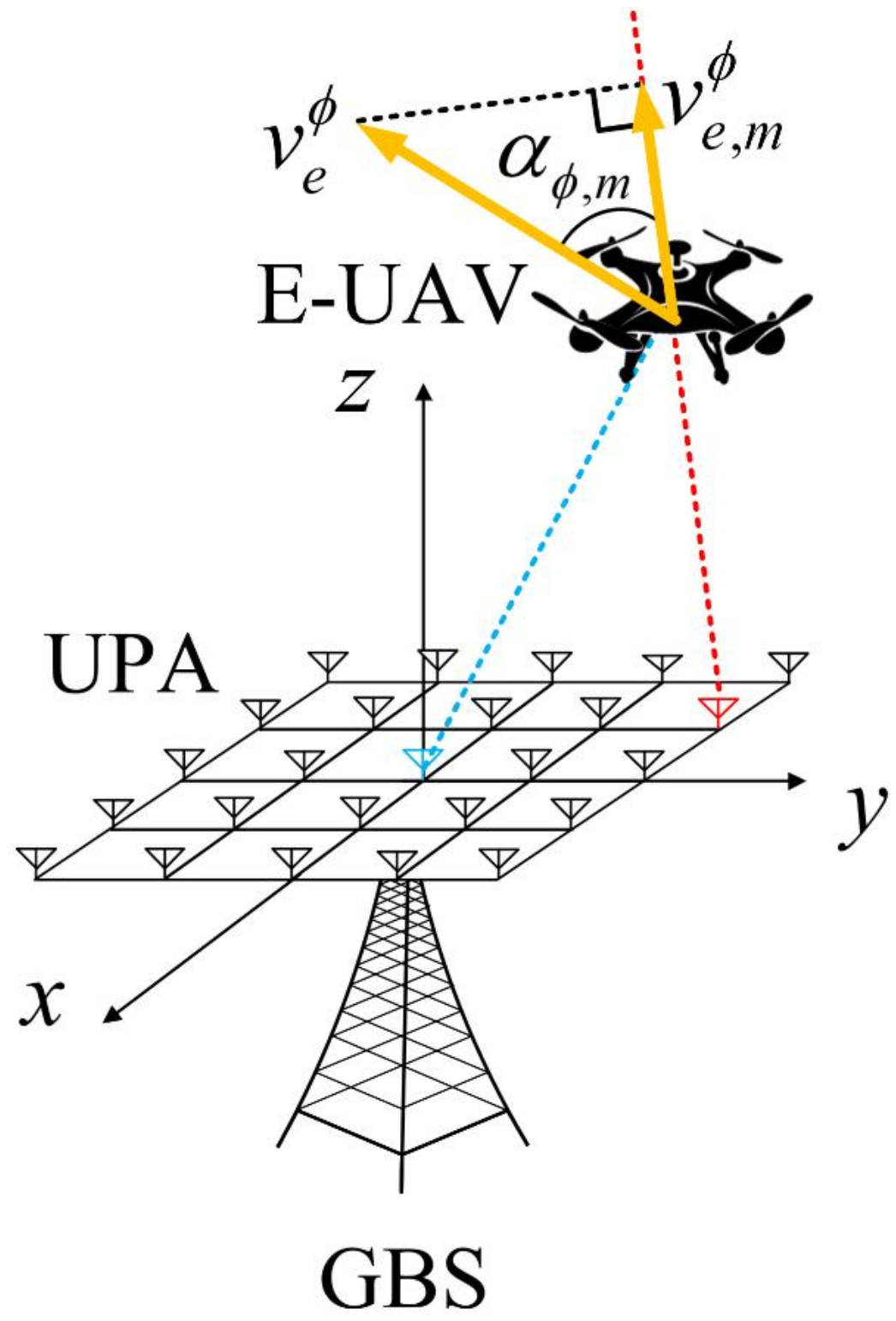}
		\caption{}
		\label{projection_3}%文中引用该图片代号
	\end{subfigure}
	\caption{Illustration of the geometrical relationship between $\tilde{\mathbf{v}}_{e,l}$ and $v_{l,m}^e$.}
	\label{projection_v}
\end{figure*}

{Thanks to the various values of $f_{l,m}^{\text{D}}, \forall m$, the ML method~\cite{NFC_V} can be utilized for the estimation of $\tilde{\mathbf{v}}_{e,l}$ based on the relationship between $\mathbf{f}_l^D$ and $\tilde{\mathbf{v}}_{e,l}$ as shown in Eq.~\eqref{eq:doppler-calculation}.}
Specifically, the ML optimization problem is formulated as follows: 
\begin{equation}
\label{eq:origingal problem}
\left(\hat{\beta},\hat{{\mathbf{v}}}_{e,l}\right)=\arg\min_{\beta,\tilde{\mathbf{v}}_{e,l}}\left\|\mathbf{Y}^{b}_l-\beta\mathbf{X}_l\left(\tilde{\mathbf{v}}_{e,l} \right)\right\|_F^2,
\end{equation}
where $\hat{\beta}$ and $\hat{\mathbf{v}}_{e,l}$ denote the estimated values of $\beta$ and $\tilde{\mathbf{v}}_{e,l}$, respectively. 
For given $\tilde{\mathbf{v}}_{e,l}$, the optimal $\hat{\beta}$ is given by 
\begin{equation}
\label{eq:beta }
\hat{\beta}=\frac{\operatorname{tr}\left(\mathbf{Y}^{b}_l\mathbf{X}_l^H\left(\tilde{\mathbf{v}}_{e,l} \right)\right)}{\|\mathbf{X}_l\left(\tilde{\mathbf{v}}_{e,l}  \right)\|_{F}^{2}}.
\end{equation}
% \vspace{-0.4cm}
Substituting the estimated $\hat{\beta}$ into the original problem $\eqref{eq:origingal problem}$, we obtain the following optimization subproblem for solving $\tilde{\mathbf{v}}_{e,l} $:
\begin{equation}
\label{eq:original-problem-2}
    \hat{\mathbf{v}}_{e,l} =\arg\max_{\tilde{\mathbf{v}}_{e,l} }g\left(\tilde{\mathbf{v}}_{e,l}\right)=\frac{\left|\mathrm{tr}\left(\mathbf{Y}^{b}_l\mathbf{X}_l^H\left(\tilde{\mathbf{v}}_{e,l}\right)\right)\right|^2}{\|\mathbf{X}_l\left(\tilde{\mathbf{v}}_{e,l}\right)\|_F^2}.
\end{equation}
Since problem~\eqref{eq:original-problem-2} is highly non-convex w.r.t. $\tilde{\mathbf{v}}_{e,l}$, the gradient descent algorithm can be utilized to find the sub-optimal solution~\cite{NFC_V}. Specifically, according to the chain rule, the gradient of $g\left(\tilde{\mathbf{v}}_{e,l}  \right)$ w.r.t. ${v}^i_{e,l}, \forall i\in\{r,\theta,\phi\}$ is given by
\begin{equation}
\label{eq:chain rule}
    \frac{\partial g\left(\tilde{\mathbf{v}}_{e,l}  \right)}{\partial v^{i}_{e,l}}=2\mathrm{Re}\left\{\mathrm{tr}\left(\frac{\partial g\left(\tilde{\mathbf{v}}_{e,l}  \right)}{\partial\mathbf{X}_l^{T}\left(\tilde{\mathbf{v}}_{e,l}\right)}\frac{\partial\mathbf{X}_l\left(\tilde{\mathbf{v}}_{e,l}\right)}{\partial v^i_{e,l}}\right)\right\},
\end{equation}
where
\begin{align}
   \label{eq:chain 1}
    \frac{\partial g\left(\tilde{\mathbf{v}}_{e,l}  \right)}{\partial\mathbf{X}_l^T\left(\tilde{\mathbf{v}}_{e,l}\right)}=& \frac{\mathrm{tr}\left(\mathbf{Y}^{b}_l\mathbf{X}_l^H\left(\tilde{\mathbf{v}}_{e,l}\right)\right)\left(\mathbf{Y}^{b}_l\right)^H}{\|\mathbf{X}_l\left(\tilde{\mathbf{v}}_{e,l}\right)\|_F^2}\notag\\&-\frac{\left|\mathrm{tr}\left(\mathbf{Y}^{b}_l\mathbf{X}_l^H\left(\tilde{\mathbf{v}}_{e,l}\right)\right)\right|^2\mathbf{X}_l^H\left(\tilde{\mathbf{v}}_{e,l}\right)}{\|\mathbf{X}_l\left(\tilde{\mathbf{v}}_{e,l}\right)\|_F^4}. 
\end{align}
Then, the next step is to calculate the following partial derivations:

\begin{subequations}
 \begin{equation}   
\begin{aligned}
\label{eq:chain 2}
    &\frac{\partial\mathbf{X}_l\left(\tilde{\mathbf{v}}_{e,l}\right)}{\partial v^i_{e,l}} =\left[\frac{\partial\mathbf{G}_l\left(1\right)}{\partial v^i_{e,l}}\mathbf{x}_l(1),\ldots,\frac{\partial\mathbf{G}_l\left(N\right)}{\partial v^i_{e,l}}\mathbf{x}_l(N)\right],
\end{aligned}
\end{equation}
\begin{equation}
\begin{aligned}
\label{eq:H gradiant}
    &\frac{\partial\mathbf{G}_l\left(n\right)}{\partial v^i_{e,l}}=2\mathbf{A}(r^e_l,\theta^e_l,\phi^e_l)\odot\left(\frac{\partial\mathbf{d}_{l,n}\left(\tilde{\mathbf{v}}_{e,l}\right)}{\partial v^i_{e,l}}\mathbf{d}_{l,n}^T\left(\tilde{\mathbf{v}}_{e,l}\right)\right),
\end{aligned}
\end{equation}
\begin{equation}
 \begin{aligned}
\label{eq:d gradiant}
    &\frac{\partial\mathbf{d}_{l,n}\left(\tilde{\mathbf{v}}_{e,l}\right)}{\partial v^i_{e,l}}=-j\frac{2\pi}{\lambda}nT_s\\
& \times\left[d_{l,n,1}\left(\tilde{\mathbf{v}}_{e,l}\right)\frac{\partial v_{e,l,1}}{\partial v^i_{e,l}},\ldots,d_{l,n,M}\left(\tilde{\mathbf{v}}_{e,l}\right)\frac{\partial v_{e,l,M}}{\partial v^i_{e,l}}\right]^T,
\vphantom{\int}
\end{aligned}
\end{equation}
\end{subequations}
where $d_{l,n,m}\left(\tilde{\mathbf{v}}_{e,l}\right)$ denotes the $m$-th element of $\mathbf{d}_{l,n}\left(\tilde{\mathbf{v}}_{e,l}\right)$. According to \eqref{eq:v-3D}, the partial derivation $\frac{\partial v_{e,l,m}}{\partial v^i_{e,l}}$ in $\eqref{eq:d gradiant}$ can be easily obtained, for which the details is omitted here. 

Finally, based on the partial derivations $\frac{\partial g\left(\tilde{\mathbf{v}}_{e,l}  \right)}{\partial v^{i}_{e,l}}, \forall i\in\left\{r,\theta,\phi\right\}$, $\tilde{\mathbf{v}}_{e,l}$ is updated through the following gradient-descent steps:  
 \begin{equation}
 \label{eq:v gradiant}
    \tilde{\mathbf{v}}_{e,l} ^{(t+1)}=\tilde{\mathbf{v}}_{e,l} ^{(t)}+\eta^{(t)}\left.\frac{\partial g\left(\tilde{\mathbf{v}}_{e,l} \right)}{\partial\tilde{\mathbf{v}}_{e,l}}\right|_{\tilde{\mathbf{v}}_{e,l} =\tilde{\mathbf{v}}_{e,l} ^{(t)}},
\end{equation}
where $\tilde{\mathbf{v}}_{e,l} ^{(t)}$ and $\eta^{(t)}$ denote the updated value of $\tilde{\mathbf{v}}_{e,l}$ and the step size in the $t$-th iteration. The details of the gradient-descent algorithm is shown in \textbf{Algorithm~\ref{alg:gradient-descent}}. {In \textbf{Algorithm~\ref{alg:gradient-descent}}, the complexity for calculating $\partial g\left(\tilde{\mathbf{v}}_{e,l}  \right)/\partial v^{i}_{e,l}$ is given by
 $\mathcal{O}\left( M^2 N\right)$. Denote $I_{\text{grad}}$ as the number of iterations for the gradient descent convergence. Then, the overall computational complexity of \textbf{Algorithm~1} is $\mathcal{O}\left(I_{\text{grad}}M^2N\right)$.}

\begin{algorithm}[tp]
    \renewcommand{\algorithmicrequire}{\textbf{Input:}}
    \renewcommand{\algorithmicensure}{\textbf{Output:}}
  \caption{{Proposed Algorithm for Optimizing $\tilde{\mathbf{v}}_{e,l}$}}
  \begin{algorithmic}[1]
  \label{alg:gradient-descent}
     \STATE Initialize feasible point $\tilde{\mathbf{v}}_{e,l}^{(0)}$
  and step size $\eta^{(0)}$.
        
      \STATE   Set iteration index $t=0$.
      \REPEAT
      \STATE Solve problem $\eqref{eq:v gradiant}$ to obtain $\tilde{\mathbf{v}}_{e,l}^{(t)}$;
            \STATE Update $\tilde{\mathbf{v}}_{e,l}=\tilde{\mathbf{v}}_{e,l}^{(t)}$;
      \STATE $t =t+1$;
      \UNTIL the fractional increment of $\tilde{\mathbf{v}}_{e,l}$ is less than the threshold $\mu_1$.
  \end{algorithmic}
\end{algorithm}

\subsection{EKF-Based E-UAV State Tracking}
To address inevitable errors of the sensed E-UAV's states, i.e., locations and velocities, obtained with the echo signals as described in Section III-A, we propose an EKF-based method to fuse the predicted states and the measured ones~\cite{weizhiqiang} for improving the E-UAV trajectory tracking accuracy. In the following, we first construct the state measurement error model and the state prediction model, and then introduce the EKF-based data fusion approach.
\label{sec:Tracking Model}
\subsubsection{State Measurement Error Model}
In the $l$-th CPI, denote the sensing errors for $r_{l}^e$, $\theta_l^e$, $\phi_l^e$, $v_{e,l}^r$, $v_{e,l}^{\theta}$, and $v_{e,l}^{\phi}$ as Gaussian random variables $z_{r^e_{l}}$, $z_{\theta^e_{l}}$, $z_{\phi^e_{l}}$, $z_{v^{r}_{e,l}}$, $z_{v^{\theta}_{e,l}}$ and $z_{v^{\phi}_{e,l}}$, respectively, each with zero mean and variances $\sigma^2_{r^e_{l}}$, $\sigma^2_{\theta^e_{l}}$, $\sigma^2_{\phi^e_{l}}$, $\sigma^2_{v^r_{e,l}}$, $\sigma^2_{v^{\theta}_{e,l}}$ and $\sigma^2_{v^{\phi}_{e,l}}$, respectively. Then, the measured states can be expressed as
\begin{subequations}
\label{eq:measure model}
\begin{equation}
\label{eq:measure model_1}
\begin{aligned}
&\hat{r}^e_{l} =\|\mathbf{q}_ {e,l}\|+z_{r^e_{l}},  
\end{aligned}
\end{equation}
\begin{equation}
\label{eq:measure model_2}
\begin{aligned}
&\hat{\theta}^e_{l} =\arcsin{\frac{x^e_{l}}{\sqrt{\left|x^e_{l}\right|^2+\left|y^e_{l}\right|^2}}}+z_{\theta^e_{l}},  
\end{aligned}
\end{equation}
\begin{equation}
\label{eq:measure model_3}
\begin{aligned}
&\hat{\phi}^e_{l} =\arcsin{\frac{z^e_{l}}{\|\mathbf{q}_{e,l}\|}}+z_{\phi^e_ {l}}, 
\end{aligned}
\end{equation}
\begin{equation}
\label{eq:measure model_4}
\begin{aligned}
&\hat{v}^r_{e,l} =\frac{v^x_{e,l}x^e_{l}+v^y_{e,l}y^e_ {l}+v^z_{e,l}z^e_ {l}}{\|\mathbf{q}_{e,l}\|}+z_{v^r_{e,l}}, 
\end{aligned}
\end{equation}
\begin{equation}
\label{eq:measure model_5}
\begin{aligned}
&\hat{v}^{\theta}_{e,l} = \frac{v^x_{e,l}y^e_{l}-v^y_{e,l}x^e_{l}}{\sqrt{\left|x^e_{l}\right|^2+\left|y^e_{l}\right|^2}}+z_{v^{\theta}_{e,l}}, 
\end{aligned}
\end{equation}
\begin{equation}
\label{eq:measure model_6}
\begin{aligned}
&\hat{v}^{\phi}_{e,l} = \frac{{v}^z_{e,l} \sqrt{\left|x^e_ {l}\right|^2+\left|y^e_ {l}\right|^2}}{\|\mathbf{q}_{e,l}\|}\\  
&+\frac{v^x_{e,l} x^e_{l} + v^y_{e,l}y^e_{l}}{\sqrt{\left|x^e_{l}\right|^2+\left|y^e_{l}\right|^2}} \times \frac{z^e_{l}}{\|\mathbf{q}_{e,l}\|} +z_{v^{\phi}_{e,l}}.
\end{aligned}
\end{equation}  
\end{subequations}

Assuming that the measurement errors for each state are uncorrelated, we can construct the measurement error covariance matrix as $\mathbf{P}_l=\text{diag}\{\sigma^2_{r^e_{l}}, \sigma^2_{{\theta}^e_{l}}, \sigma^2_{{\phi}^e_{l}}, \sigma^2_{v^r_{e,l}}, \sigma^2_{v^{\theta}_{e,l}}, \sigma^2_{v^{\phi}_{e,l}}\}$. It has been demonstrated in~\cite{celiangwucha} that the measurement errors are inversely proportional to the signal-to-noise ratio (SNR) of the echo signals. Therefore, we have $\sigma^2_{r^e_{l}}=\frac{c_{r}}{\mathrm{SNR}_l} $,  $\sigma^2_{\theta^e_{l}}=\frac{c_{{\theta}}}{\mathrm{SNR}_l} $,  $\sigma^2_{\phi^e_{l}}=\frac{c_{{\phi}}}{\mathrm{SNR}_l} $, $\sigma^2_{v_ {r,l}}=\frac{c_{v_ {r}}}{\mathrm{SNR}_l}$ , $\sigma^2_{v_ {\theta}}=\frac{c_{v_ {\theta}}}{\mathrm{SNR}_l}$ and $\sigma^2_{v_ {\phi,l}}=\frac{c_{v_ {\phi}}}{\mathrm{SNR}_l}$,
where $c_r$, $c_{{\theta}}$, $c_{{\phi}}$, $c_{v_{r}}$,$c_{v_{\theta}}$ and $c_{v_{\phi}}$ are the scaling factors~\cite{snr_c}, while $\mathrm{SNR}_l$ is given by 
\begin{equation}
\label{eq:snr}
\mathrm{SNR}_l=\frac{\mathrm{\sigma_{RCS}}{\beta_0}^2MN|\boldsymbol{\alpha}(r^{e}_{l}, \theta^{e}_{l}, \phi^{e}_{l})\mathbf{w}^{e}_{l}|^2}{4\pi\sigma_b^2\left(r^{e}_{l}\right)^4}.
\end{equation}
Denote the E-UAV's ground-truth state vector under the Cartesian coordinate system and the measured state vector under the spherical coordinate system as $\mathbf{s}^e_ {l}=\left[x^e_{l},y^e_{l},z^e_{l},v^x_{e,l},v^y_{e,l},v^z_{e,l}\right]^T$ and
$\hat{\mathbf{u}}_{l}=\left[\hat{r}^e_{l},\hat{\theta}^e_{l},\hat{\phi}^e_ {l},\hat{v}^r_{e,l},\hat{v}^{\theta}_{e,l},\hat{v}^{\phi}_{e,l} \right]^T$, respectively.
Then, \eqref{eq:measure model} can be rewritten as
\begin{equation}
\label{eq:measure model vector}
   \hat{\mathbf{u}}_{l}=\mathbf{\chi}_{l}(\mathbf{s}^e_{l})+\mathbf{z}_{l},
\end{equation}
 where $\mathbf{z}_{l}=[z_{r^e_{l}}, z_{\theta^e_{l}}, z_{\phi^e_{l}}, z_{v^r_{e,l}}, z_{v^{\theta}_{e,l}}, z_{v^{\phi}_{e,l}}]^T$ denotes the noise vector, and the function $\chi(\cdot)$ can be derived based on
$\eqref{eq:measure model}$ . 
\subsubsection{State Prediction Model}
To accurately characterize the temporal motion of the E-UAV, it is essential to establish a state prediction model. Specifically, the state prediction equation can be expressed as
\begin{equation}
\label{eq:measure model vector}
   \mathbf{s}^e_{l}=\mathbf{F}\mathbf{s}^e_{l-1}+\boldsymbol{\zeta}_ {l},
\end{equation}
where $\boldsymbol{\zeta}
_l=\left[\zeta_{x^{e}_l}, \zeta_{y^e_l}, \zeta_{z^e_l}, \zeta_{v^x_{e,l}}, \zeta_{v^y_{e,l}},\zeta_{v^z_{e,l}}\right]^T$, $\zeta_{x^e_l}, \zeta_{y^e_l}, \zeta_{z^e_l}, \zeta_{v^x_{e,l}}, \zeta_{v^y_{e,l}}$, and $\zeta_{v^z_{e,l}}$  denote the
prediction errors for $x^e_{l},y^e_{l},z^e_{l},v^x_{e,l},v^y_{e,l}
$, and $v^z_{e,l}$ respectively, with $\zeta_{x^e_l}\sim \mathcal{N} (0,\sigma_{x^e_l }^{2}), \zeta_{y^e_l}\sim \mathcal{N} (0,\sigma_{y^e_l }^{2}), \zeta_{z^e_l}\sim \mathcal{N} (0,\sigma_{z^e_l }^{2}), \zeta_{v^x_{e,l}}\sim \mathcal{N} (0,\sigma_{v^x_{e,l}}^{2}), \zeta_{v^y_{e,l}}\sim \mathcal{N} (0,\sigma_{v^y_{e,l}}^{2})$ and $\zeta_{{v^z_{e,l}}}\sim \mathcal{N} (0,\sigma_{v^z_{e,l}}^{2})$, respectively. Again, assuming mutually independent state prediction errors, we have $\boldsymbol{\zeta}_l\sim\mathcal{N}(0,\mathbf{Q}_{s})$ with $\mathbf{Q}_{s}=\mathrm{diag}\{\sigma_{x^e_l }^{2},\sigma_{y^e_l }^{2},\sigma_{z^e_l }^{2},\sigma_{v^x_{e,l}}^{2},\sigma_{v^y_{e,l} }^{2},\sigma_{v^z_{e,l} }^{2}\}$. Moreover, $\mathbf{F}\in\mathbb{R}^{6\times6}$ is the state transition matrix. In the prediction model, we make the assumption of constant-velocity movement between adjacent CPIs and express the state transition matrix accordingly as 
\begin{equation}
\label{eq:state transition matrix}
\left.\mathbf{F}=\left[\begin{array}{cc}\mathbf{I}_{3}&\Delta t\mathbf{I}_{3}\\\mathbf{0}_{3}&\mathbf{I}_{3}\end{array}\right.\right].
\end{equation}

\subsubsection{EKF-Based Data Fusion} 
Due to the nonlinearity of~\eqref{eq:measure model}, we adopt the EKF for fusing the E-UAV's measured states and the predicted ones~\cite{EKF}. In the $l\text{-}\mathrm{th}$ CPI, we assume that the GBS has the E-UAV's state estimation $\hat{\boldsymbol{\mathbf{s}}}^e_{l-1}$ from the $(l-1)$-th CPI, along with the corresponding covariance matrix $\mathbf{C}_{l-1}$ of the estimation errors. Then, the state prediction equation and the corresponding prediction error covariance matrix are respectively given by 
\begin{equation}
\label{eq: state prediction equation}
\hat{\mathbf{s}}^e_{l|l-1}=\mathbf{F}\hat{\mathbf{s}}^e_{l-1},
\end{equation}
\begin{equation}
\label{eq: corresponding prediction covariance matrix}
\mathbf{C}_{l|l-1}=\mathbf{F}\mathbf{C}_{l-1}\mathbf{F}^T+\mathbf{Q}_{s},
\end{equation}
where $\hat{\mathbf{s}}^e_{l|l-1}$ denotes the predicted state of the E-UAV in the $l\text{-}\mathrm{th}$ CPI. With the predicted state $\hat{\mathbf{s}}^e_{l|l-1}$ and the measured one $\hat{\mathbf{u}}_l$, the E-UAV's state in the $l\text{-}\mathrm{th}$ CPI is updated by 
\begin{equation}
\label{eq:Kalman gain}
\hat{\mathbf{s}}^e_{l}=\mathbf{\hat{s}}^e_{l|l-1}+\mathbf{K}_{l}(\mathbf{u}_l-\mathbf{h}(\mathbf{\hat{s}}^e_{l|l-1})),
\end{equation}
where $\mathbf{K}_{l}$ denotes the Kalman gain that is given by
\begin{equation}
\label{eq:Kalman gain}
\mathbf{K}_{l}=\mathbf{C}_{l|l-1}\boldsymbol{\Sigma_l}^T\left(\boldsymbol{\Sigma}_l\mathbf{C}_{l|l-1}\boldsymbol{\Sigma}_l^T+\mathbf{P}_ {l}\right)^{-1}.
\end{equation}
The covariance matrix of the estimation error, i.e., $\hat{\mathbf{s}}_l^e-s_l^e$, is given by~\cite{weizhiqiang}

\begin{equation}
\label{eq:the covariance matrix gai}
\mathbf{C}_l=
\left(\mathbf{C}_{l|l-1}^{-1}+\boldsymbol{\Sigma}_l^H\mathbf{P}_l^{-1}\boldsymbol{\Sigma}_l\right)^{-1},
\end{equation}
where $\Sigma_l=\frac{ \partial \chi_l }{ \partial \mathbf{s}^{e}_{l}}|_{\mathbf{s}^{e}_{l}=\hat{\mathbf{s}}^e_{l|l-1}}$ denotes the Jacobian matrix of $\mathbf{h}(\cdot)$ w.r.t. $\mathbf{s}^{e}_{l}$.
It is worth noting that $\mathbf{C}_l$ is the key variable to evaluate the E-UAV trajectory tracking accuracy, which highly depends on the SNR of the echo signals received at the GBS due to $\mathbf{P}_l$. Therefore, the value of $\mathbf{C}_l$ will be taken into consideration for the joint GBS beamforming and C-UAV trajectory design in the next section. {The overall E-UAV trajectory tracking workflow is shown in Fig.~\ref{fig:tracking_in_CPIs}.} 
\begin{figure}
    \centering
\includegraphics[width=0.8\linewidth]{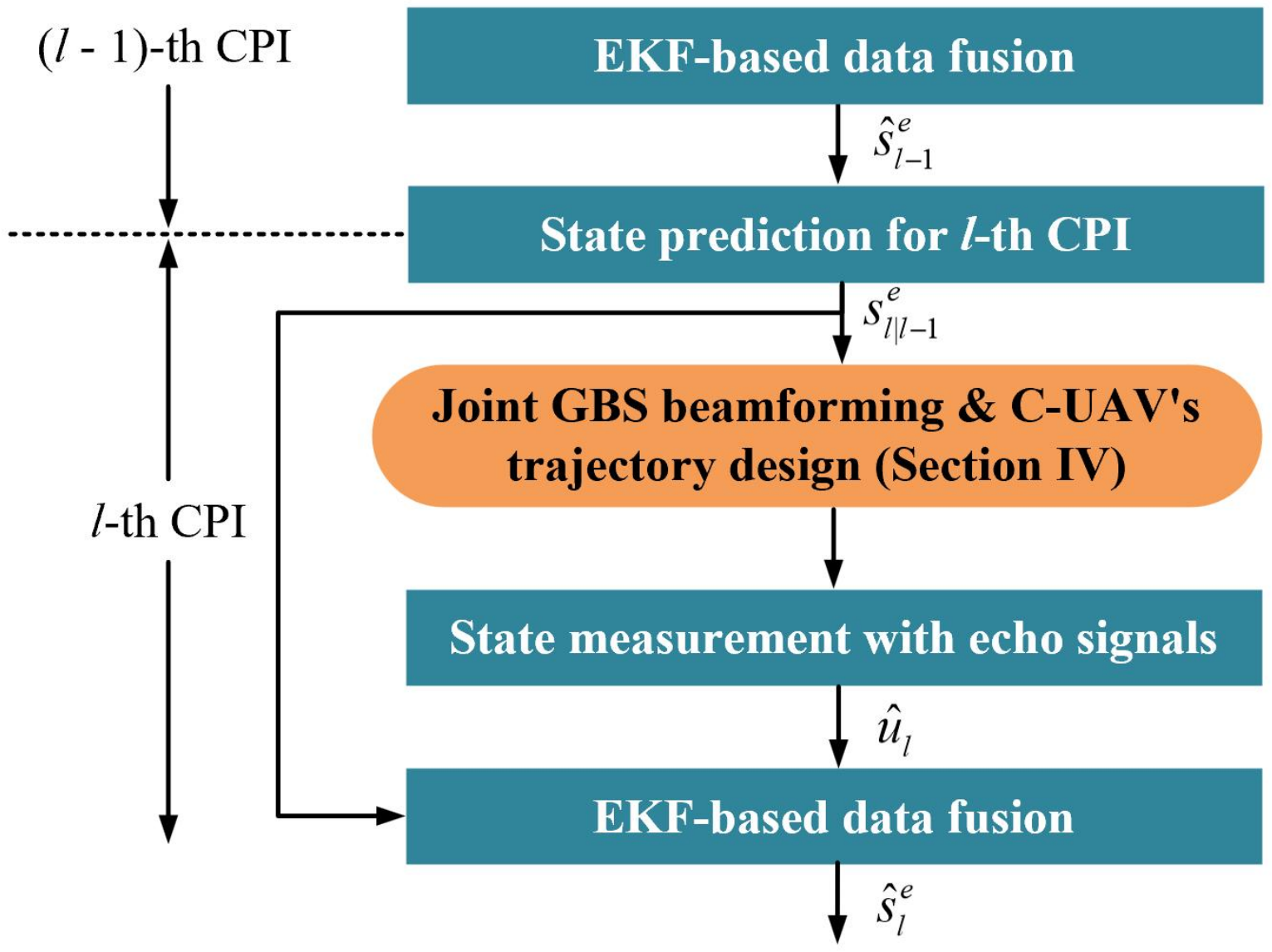}
    \caption{{The overall E-UAV trajectory tracking workflow.}}
    \label{fig:tracking_in_CPIs}
\end{figure}

\section{Online Joint GBS Beamforming and C-UAV Trajectory Design Scheme}
{With the E-UAV trajectory tracking scheme proposed in Section III, the GBS can obtain the real-time E-UAV's locations, and thus facilitates the online GBS beamforming and C-UAV trajectory design. On this basis, we formulate the optimization problem with the aim of maximizing the secrecy rate for the $l$-th CPI as follows:}
\begin{subequations}
\label{eq:optimization_problem}
\begin{equation}
\label{eq:objective_function}
    \max_{\mathbf{q}_{c,l},\mathbf{w}^{c}_{l},\mathbf{w}^{e}_{l}}\hat{R}^{s}_{l},
\end{equation}
% \vspace{-0.4cm}
\begin{equation}
\label{eq:v}
   {\rm{s.t.}} \ \ \|\mathbf{q}_{c,l}-\mathbf{q}_{c,l-1}\|\leq V_\text{max}\Delta t,
\end{equation}
% \vspace{-0.4cm}
\begin{equation}
\label{eq:field}
  \mathbf{q}_{\mathrm{min}}\leq\mathbf{q}_{c,l}\leq\mathbf{q}_{\mathrm{max}},
\end{equation}
% \vspace{-0.4cm}
\begin{equation}
\label{eq:collision}
  ||\mathbf{q}_{c,l}-\hat{\mathbf{q}}_{e,l|l-1}||^2\geq d_{\min}^2,
\end{equation}
\begin{equation}
\label{eq:power}
\left\|\mathbf{w}^{c}_{l}\right\|^2+\left\|\mathbf{w}^{e}_{l}\right\|^2\leq P_\mathrm{max},
\end{equation}
% \vspace{-0.4cm}
\begin{equation}
\label{eq:MSE constraint 1}
\mathrm{tr}(\mathbf{C}_{l})\leq  \Gamma.
\end{equation}
% \vspace{-0.4cm}
\end{subequations}
In~\eqref{eq:optimization_problem}, $\hat{R}^{s}_{l}$ denotes the secrecy rate calculated with the predicted E-UAV's location $\hat{\mathbf{q}}_{e,l|l-1}$, for which the full expression will be provided later. \eqref{eq:v} restricts that the C-UAV's maximum speed does not exceed $V_{\mathrm{max}}$. {\eqref{eq:field} confines the C-UAV's flight within the 3D space $\left[\mathbf{q}_{\min},\mathbf{q}_{\max}\right]$}. \eqref{eq:collision} is the collision avoidance constraint, where $d_{\min}$ is the minimum safe distance between the C-UAV and the E-UAV.  \eqref{eq:power} guarantees that the GBS transmit power does not exceed $P_{\text{max}}$. {$\eqref{eq:MSE constraint 1}$ ensures that the mean squared error (MSE) of the E-UAV's state estimation is not larger than the threshold $\Gamma$.} The optimization problem~\eqref{eq:optimization_problem} is intractable to solve as it is non-convex w.r.t. $\mathbf{q}_{c,l},\mathbf{w}^{c}_{l},\mathbf{w}^{e}_{l}$, and the involved variables are highly coupled. Therefore, we propose to invoke the AO algorithm to decouple the original problem into the GBS beamforming and the C-UAV trajectory design subproblems, where each subproblem is iteratively optimized while keeping the other one fixed.

% Based on the CSI of the E-UAV, we can design the trajectory and the precoding vector to solve the problem.
% However, the problem is still difficult to be solved optimally due to
% the non-concave objective function and constraints. Therefore, considering the highly coupling of the beamforming
% variables and highly difficulty of directly optimizing the C-UAV
% trajectory, we propose an alternating optimization algorithm
% by alternately optimizing the communication and sensing
% beamforming as well as the UAV trajectory to find the
% suboptimal solution.
% With the previous state of the E-UAV, the predicted state information for the $l\text{-th}$ CPI can be obtained through the state update equation.
% The precoding vectors $\mathbf{w}^{c}_{l}$ and $\mathbf{w}^{e}_{l}$ can be derived from the eigenvectors of $\mathbf{W}_{c,l}$ and $\mathbf{W}_{e,l}$ corresponding to their maximum eigenvalues, respectively. Additionally, we define $\mathbf{H}^{bc}_{l}=\mathbf{h}^{bc}_{l}\left(\mathbf{h}^{bc}_{l}\right)^H$. The optimization problem is based on the predicted state of the E-UAV. Thus, in the subsequent analysis, $\hat{\mathbf{H}}^{be}_{l|l-1}$ is utilized to denote $\mathbf{H}^{be}_{l}$, and $\hat{\mathbf{q}}_{e,l|l-1}$ is utilized to denote $\mathbf{q}_{e,l}$. 

%\end{itemize}
\subsection{GBS Beamforming Optimization}
To start with, we first define $\mathbf{W}_{c,l}=\mathbf{w}^{c}_{l}\left(\mathbf{w}^{c}_{l}\right)^H$ and $\mathbf{W}_{e,l}=\mathbf{w}^{e}_{l}\left(\mathbf{w}^{e}_{l}\right)^H$. Then, the GBS beamforming subproblem can be formulated as
\begin{subequations}
\label{eq:beamforming-subproblem}
\begin{equation}
\label{eq:P2}
\max_{\mathbf{W}_{c,l},\mathbf{W}_{e,l}} \hat{R}^{s}_{l}
\end{equation}
\begin{equation}
\label{eq:power 2}
  {\rm{s.t.}} \ \ \mathrm{tr}(\mathbf{W}_{c,l})+\mathrm{tr}(\mathbf{W}_{e,l})\leq P_\mathrm{max},
\end{equation}
\begin{equation}
\label{eq:se_constraint}
\mathbf{W}_{c,l}\succeq 0,\mathbf{W}_{e,l}\succeq 0,
\end{equation}
\begin{equation}
\label{eq:r}
\mathrm{Rank}\left(\mathbf{W}_{c,l}\right)=1,\mathrm{Rank}\left(\mathbf{W}_{e,l}\right)=1,
\end{equation}
\begin{equation}
\label{eq:mse-constraint}
\eqref{eq:MSE constraint 1}.
\end{equation}
\end{subequations}
In {~\eqref{eq:P2}}, $\hat{R}^{s}_{l}$ is given by
\begin{align}
\label{eq:objectory function}
    \hat{R}^{s}_{l}=&\left[\log_2\left(1+\frac{\mathrm{tr}(\mathbf{H}^{bc}_{l}\mathbf{W}_{c,l})}{\mathrm{tr}(\mathbf{H}^{bc}_{l}\mathbf{W}_{e,l})+\sigma_\mathrm{c}^2}\right)\right.\notag&\\
    &\left.-\log_2\left(1+\frac{\mathrm{tr}(\hat{\mathbf{H}}^{be}_{l|l-1}\mathbf{W}_{c,l})}{\mathrm{tr}(\hat{\mathbf{H}}^{be}_{l|l-1}\mathbf{W}_{e,l})+\sigma_\mathrm{e}^2}\right)\right]^{+},
\end{align}
where $\mathbf{H}^{bc}_{l}=\mathbf{h}^{bc}_{l}\left(\mathbf{h}^{bc}_{l}\right)^H$ and $\hat{\mathbf{H}}^{be}_{l|l-1}=\hat{\mathbf{h}}^{be}_{l|l-1}\left(\hat{\mathbf{h}}^{be}_{l|l-1}\right)^H$ with $\hat{\mathbf{h}}^{be}_{l|l-1}$ denoting the channel between the GBS and the E-UAV calculated by the predicted E-UAV's location $\hat{\mathbf{q}}_{e,l|l-1}$. To facilitate the design, we introduce the auxiliary exponential variables $\tau$, $\varepsilon$, $\delta$, and $\epsilon$, and transform problem~\eqref{eq:beamforming-subproblem} as
\begin{subequations}
\label{eq:beamforming-optimization-transform}
    \begin{equation}
\label{eq:P2}
\max_{\mathbf{W}_{c,l},\mathbf{W}_{e,l},\tau,\varepsilon,\delta,\epsilon} \tau-\varepsilon-\delta+\epsilon
\end{equation}
\begin{equation}
\label{eq:auxiliary constraint 1}
e^{\tau}\leq\mathrm{tr}\Big(\mathbf{H}^{bc}_{l}\mathbf{W}_{e,l}\Big)+\mathrm{tr}\Big(\mathbf{H}^{bc}_{l}\mathbf{W}_{c,l}\Big)+\sigma_{c}^2,
\end{equation}
\begin{equation}
\label{eq:auxiliary constraint 2}
e^{\varepsilon}\geq\mathrm{tr}\big(\mathbf{H}^{bc}_{l}\mathbf{W}_{e,l}\big)+\sigma_{c}^2,
\end{equation}
\begin{equation}
\label{eq:auxiliary constraint 3}
e^{\delta}\geq\mathrm{tr}\Big(\hat{\mathbf{H}}^{be}_{l|l-1}\mathbf{W}_{c,l}\Big)+\mathrm{tr}\Big(\hat{\mathbf{H}}^{be}_{l|l-1}\mathbf{W}_{e,l}\Big)+\sigma_{e}^2,
\end{equation}
\begin{equation}
\label{eq:auxiliary constraint 4}
e^{\epsilon}\leq\mathrm{tr}\Big(\hat{\mathbf{H}}^{be}_{l|l-1}\mathbf{W}_{e,l}\Big)+\sigma_{e}^2,
\end{equation}
\begin{equation}
    \eqref{eq:power 2}-\eqref{eq:r}, \eqref{eq:MSE constraint 1}.
\end{equation}
\end{subequations}
Problem~\eqref{eq:beamforming-optimization-transform} is still non-convex due to constraints $\eqref{eq:auxiliary constraint 2}$, $\eqref{eq:auxiliary constraint 3}$, \eqref{eq:r}, and \eqref{eq:mse-constraint}. For $\eqref{eq:auxiliary constraint 2}$ and $\eqref{eq:auxiliary constraint 3}$, we apply the first-order Taylor expansion to obtain the following two relaxed constraints:
\begin{equation}
\label{eq:auxiliary constraint 2 gai}
\mathrm{tr}\big(\mathbf{H}^{bc}_{l}\mathbf{W}_{e,l}\big)+\sigma_{c}^2\leq e^{\varepsilon^{\left(j\right)}}(\varepsilon-\varepsilon^{\left(j\right)}+1),
\end{equation}
\begin{equation}
\begin{aligned}
\label{eq:auxiliary constraint 3 gai}
\mathrm{tr}\big(\hat{\mathbf{H}}^{be}_{l|l-1}\mathbf{W}_{c,l}\big)+&\mathrm{tr}\big(\hat{\mathbf{H}}^{be}_{l|l-1}\mathbf{W}_{e,l}\big)+\sigma_{e}^{2}\\&\leq e^{\delta^{\left(j\right)}}(\delta-\delta^{\left(j\right)}+1), 
\end{aligned}
\end{equation}
where $\varepsilon^{(j)}$ and $\delta^{(j)}$ are the given local points of $\varepsilon$ and $\delta$ in the $j$-th iteration of the SCA, respectively. Moreover, to deal with the non-convexity of the covariance matrix in $\eqref{eq:MSE constraint 1}$,  we introduce auxiliary variables $\begin{matrix}u_{m}\geq0,\forall m\in{\{1,...,6\}}\end{matrix}$ to upper bound the diagonal elements in $\mathbf{C}_l$, and transform constraint $\eqref{eq:MSE constraint 1}$ to the following two constraints: 
\begin{equation}
\label{eq:covariance constraint 1}
[\mathbf{C}_{l}]_{m,m}\leq u_{m}, \forall m,
\end{equation}
\begin{equation}
\label{eq:covariance constraint 2}
\sum_{i=1}^6u_m\leq \Gamma, \forall m,
\end{equation}
where $[\mathbf{C}_{l}]_{m,m}$ denotes the $m$-th diagonal element of $\mathbf{C}_{l}$. To further simplify the non-convex form of \eqref{eq:covariance constraint 1}, we utilize the Schur complement~\cite{Schur} to obtain the following equivalent constraint:
\begin{equation}
\label{eq:covariance constraint 1 gai}
\begin{bmatrix}\mathbf{C}_{l}^{-1}&\mathbf{e}_{m}\\\mathbf{e}_{m}^{T}&u_{m}\end{bmatrix}\succeq\mathbf{0},\forall m,
\end{equation}
{where $\mathbf{e}_m \in \mathbb{R}^{6\times1}$ denotes a vector whose $m$-th element is $1$ and other elements are $0$}.

Now, the only non-convexity left in {problem~\eqref{eq:beamforming-optimization-transform}} is the
rank-one constraint $\eqref{eq:r}$. To deal with this issue, we provide an equivalent difference of convex (DC) representation as
\begin{equation}
\label{eq:DC}
  \psi_{i,l}\triangleq\|\mathbf{W}_{i,l}\|_*-\|\mathbf{W}_{i,l}\|_2=0, i\in\{c,e\},
\end{equation}
where $\|\mathbf{W}_{i,l}\|_{*}$ and $\|\mathbf{W}_{i,l}\|_{2}$ denote the nuclear norm and the largest spectral norm of $\mathbf{W}_{i,l}$, respectively. Note that for any $\mathbf{W}_{i,l}$, the inequality $\|\mathbf{W}_{i,l}\|_*-\|\mathbf{W}_{i,l}\|_2\geq0$ always hold, with equality occurring if and only if
$\mathbf{W}_{i,l}$ is a rank-one matrix. Hence, the constraint \eqref{eq:r} is satisfied exclusively when $\mathbf{W}_{i,l}$ is rank-one. 
Next, we apply the penalty-based method~\cite{penalty} to tackle the constraint~\eqref{eq:DC}. Specifically, by adding \eqref{eq:DC} into the objective function as a penalty term, problem \eqref{eq:beamforming-optimization-transform} can be converted to:
\begin{subequations}
\label{eq:beamforming_problem_penalty} 
\begin{equation}
\begin{aligned}
\label{eq:beamforming_problem_penalty_obj} &\min_{\mathbf{W}_{c,l},\mathbf{W}_{e,l}} - (\tau-\varepsilon-\delta+\epsilon)+\omega\sum_{i\in\{c,e\}} \psi_{i,l},
%(\|\mathbf{W}_{i,l}\|_*-\|\mathbf{W}_{i,l}\|_2),
\end{aligned}
\end{equation}
\begin{equation}
\label{eq:else 2}
  {\rm{s.t.}} \ \  \eqref{eq:power 2}, \eqref{eq:se_constraint}, \eqref{eq:auxiliary constraint 1}, \eqref{eq:auxiliary constraint 4}, \eqref{eq:auxiliary constraint 2 gai}-\eqref{eq:auxiliary constraint 3 gai}, \eqref{eq:covariance constraint 2}, \eqref{eq:covariance constraint 1 gai},
\end{equation}
\end{subequations}
where $\omega>0$ denotes the penalty factor. As $\omega\rightarrow+\infty$ (or becomes sufficiently large), the solution $\mathbf{W}_{i,l}$ of problem \eqref{eq:beamforming_problem_penalty} will comply with the equality constraint \eqref{eq:DC}. However, if $\omega$ is initially set too large, the penalty term will dominate the objective function \eqref{eq:beamforming_problem_penalty_obj}, diminishing the influence of the desired maximum secrecy rate on the optimization result. To prevent this, we start by setting $\omega$ to a relatively small value to determine an appropriate initial point, after which $\omega$ is gradually increased until the rank-one constraint is satisfied.

To deal with the non-convexity of the penalty term in~\eqref{eq:beamforming_problem_penalty_obj}, we utilize the first-order Taylor expansion to obtain the convex upper bound in the $j$-th iteartion of the SCA as follows:
\begin{equation}
    \label{penalty_term_sca}
\|\mathbf{W}_{i,l}\|_{*}-\|\mathbf{W}_{i,l}\|_{2}\leq \mathrm{tr}\left(\mathbf{W}_{i,l}\right)-\bar{\mathbf{W}}^{(j)}_{i,l}\triangleq \psi^{(j)}_{i,l},
\end{equation}
where $\bar{\mathbf{W}}^{(j)}_{i,l}\triangleq \|\mathbf{W}_{i,l}^{\left(j\right)}\|_{2}+\mathrm{tr}\Big(\xi(\mathbf{W}_{i,l}^{\left(j\right)})\xi(\mathbf{W}_{i,l}^{\left(j\right)})^{H}(\mathbf{W}_{i,l}-\mathbf{W}_{i,l}^{\left(j\right)})\Big)$ and $\xi(\mathbf{W}_{i,l}^{\left(j\right)})$
is the eigenvector of $\mathbf{W}_{i,l}$
corresponding to the largest singular value $\sigma_{1}(\mathbf{W}_{i,l})$.
% and in the form of a DC function, we can adapt the first-order Taylor expansion to construct the upper-bound surrogate function in the $j$-th iteration of the SCA as follows:
% \textcolor{red}{
% \begin{equation}
% \label{eq:penalty SCA}
%   \begin{aligned}
%   &\|\mathbf{W}_{i,l}\|_{*}-\|\mathbf{W}_{i,l}\|_{2}\leq\|\mathbf{W}_{i,l}\|_{*}\\&-\left(\|\mathbf{W}_{i,l}^{\left(j\right)}\|_{2}+\mathrm{tr}\Big(\xi(\mathbf{W}_{i,l}^{\left(j\right)})\xi(\mathbf{W}_{i,l}^{\left(j\right)})^{H}(\mathbf{W}_{i,l}-\mathbf{W}_{i,l}^{\left(j\right)})\Big)\right),
%   \end{aligned}
% \end{equation}}
% where $\xi(\mathbf{W}_{i,l}^{\left(j\right)})$
% is the eigenvector of $\mathbf{W}_{i,l}$
% corresponding to the largest singular value $\sigma_{1}(\mathbf{W}_{i,l})$. 
Subsequently, problem \eqref{eq:beamforming-optimization-transform} can be equivalently addressed by solving the following problem:
\begin{subequations}
\label{eq:beamforming-convex}
\begin{equation}
\begin{aligned}
\label{eq:beamforming_finally}
    &\min_{\mathbf{W}_{c,l},\mathbf{W}_{e,l},\tau,\varepsilon,\delta,\epsilon} -(\tau-\varepsilon-\delta+\epsilon)+\omega\sum_{t\in\{c,e\}}\psi^{(j)}_{i,l},
\end{aligned}
\end{equation}
\begin{equation}
\label{eq:else 2}
  {\rm{s.t.}} \ \  \eqref{eq:power 2}, \eqref{eq:se_constraint}, \eqref{eq:auxiliary constraint 1}, \eqref{eq:auxiliary constraint 4}, \eqref{eq:auxiliary constraint 2 gai}-\eqref{eq:auxiliary constraint 3 gai}, \eqref{eq:covariance constraint 2}, \eqref{eq:covariance constraint 1 gai}.
\end{equation}
\end{subequations}
The relaxed problem \eqref{eq:beamforming-convex}, structured as a standard convex semi-definite program (SDP), can be effectively addressed by using optimization tools like CVX~\cite{boyd2004convex}. The specific details of the proposed algorithm for solving the original problem \eqref{eq:beamforming-subproblem} are outlined
in \textbf{Algorithm \ref{al_2}}.
Note that \textbf{Algorithm \ref{al_2}}  consists of two loops. In the inner loop, the
variable $\mathbf{W}_{c,l},\mathbf{W}_{e,l}$ are optimized by solving the relaxed convex form
\eqref{eq:beamforming-convex} iteratively. {In the outer loop, the penalty factor $\omega$ is updated with the scaling factor $c>1$ until the rank-one constraint is satisfied. }

\begin{algorithm}[tp]
    \renewcommand{\algorithmicrequire}{\textbf{Input:}}
    \renewcommand{\algorithmicensure}{\textbf{Output:}}
  \caption{Proposed Algorithm for Optimizing $\mathbf{W}_{c,l},\mathbf{W}_{e,l}$ }
  \begin{algorithmic}[1]
    \label{al_2}
       \STATE 
               Initialize feasible points $\left\{\mathbf{W}_{c,l}^{(0)}, \mathbf{W}_{e,l}^{(0)}\right\}$, and the penalty factor $\omega$.
      \REPEAT       \STATE Set iteration index $j = 0$;
         \REPEAT
          
              \STATE Solve problem $\eqref{eq:beamforming-convex}$ to obtain $\mathbf{W}_{c,l}^{(j)}$ and $\mathbf{W}_{e,l}^{(j)}$;
              \STATE Update 
              $\mathbf{W}_{c,l}=\mathbf{W}_{c,l}^{(j)},$ and $\mathbf{W}_{c,l}=\mathbf{W}_{e,l}^{(j)}$;
              \STATE $j =j+1$;
      \UNTIL the fractional decrement of the $\hat{R}_l^s$ is less than the threshold $\mu_{2}$;       \STATE Update $\omega=c\omega$;
      \UNTIL the rank-one constraint violation is below a predefined threshold $\mu_3$.

  \end{algorithmic}
\end{algorithm}

\vspace{-0.3cm}
\subsection{{C-UAV Trajectory Optimization}}
With the fixed GBS beamforming vectors $\mathbf{w}_{c,l}$ and $\mathbf{w}_{e,l}$, the C-UAV trajectory design subproblem can be formulated as
\begin{subequations}
\label{eq:P1_trajectory}
\begin{equation}
\begin{aligned}
\label{eq:P1 trajectory_obj}
    \max_{\mathbf{q}_{c,l}}  \hat{R}^{s}_{l}
\end{aligned}
\end{equation}
\begin{equation}
\label{eq:else 3}
  {\rm{s.t.}} \ \  \eqref{eq:v}- \eqref{eq:collision}.
\end{equation}
\end{subequations}
As the leakage information rate is irrelevant to the C-UAV trajectory $\mathbf{q}_{c,l}$, we can rewrite problem~\eqref{eq:P1_trajectory} as
\begin{subequations}
\label{eq:P2 trajectory}
\begin{equation}
\begin{aligned}
\label{eq:P2 trajectory_obj}
    \max_{\mathbf{q}_{c,l}}  R^{c}_{l}
\end{aligned}
\end{equation}
\begin{equation}
\label{eq:else 4}
  {\rm{s.t.}} \ \  \eqref{eq:v}- \eqref{eq:collision}.
\end{equation}
\end{subequations}
Since the steering vector $\boldsymbol{\alpha}(r^c_l,\theta^c_l,\phi^c_l)$ is highly complex w.r.t. $\mathbf{q}_{c,l}$, we propose to adopt the value of $\mathbf{q}_{c,l}$ in the $(k-1)$-th iteration of the SCA, i.e., $\mathbf{q}_{c,l}^{(k-1)}$, for approximating $\boldsymbol{\alpha}(r^c_l,\theta^c_l,\phi^c_l)$ in the $k$-th iteration~\cite{k+1_k}. As such, the communication channel $\mathbf{h}^c_l$ in the $k$-th iteration of the SCA is approximated by
\begin{equation}
\label{eq:hc_SCA}
  \mathbf{h}^{c,(k)}_{l}=\frac{\beta_0}{r^c_{l}}\boldsymbol{\alpha}\left(r^{c,(k-1)}_l,\theta^{c,(k-1)}_l,\phi^{c,(k-1)}_l\right)=\frac{\tilde{\mathbf{{h}}}^{c,(k-1)}_{l}}{r^{c}_{l}},
\end{equation}
where $\tilde{\mathbf{{h}}}^{c,(k-1)}_{l}\triangleq\beta_0\boldsymbol{\alpha}\left(r^{c,(k-1)}_l,\theta^{c,(k-1)}_l,\phi^{c,(k-1)}_l\right)$,
$r^{c,(k-1)}_l,\theta^{c,(k-1)}_l$, and $\phi^{c,(k-1)}_l$ denote the distance, azimuthal angle, and elevation angle between the C-UAV and the GBS based on $\mathbf{q}_{c,l}^{(k-1)}$. Accordingly, in the $k$-th iteration, the objective function $R_l^{c,(k)}$ is given by
\begin{equation}
\begin{aligned}
\label{eq:P3 trajectory_obj}
    &R_l^{c,(k)}= \\&\log_2\left(1+\frac{\mathrm{tr}(\mathbf{\tilde{H}}^{bc,(k-1)}_{l}\mathbf{W}_{c,l})}{\mathrm{tr}(\mathbf{\tilde{H}}^{bc,(k-1)}_{l}\mathbf{W}_{e,l})+\|\mathbf{q}_{c,l}-\mathbf{q}_b\|^2\sigma_c^2}\right),
\end{aligned}
\end{equation}
where $\mathbf{\tilde{H}}^{bc,(k-1)}_{l}=\mathbf{\tilde{h}}^{c,(k-1)}_{l}\left(\mathbf{\tilde{h}}^{c,(k-1)}_{l}\right)^H$. Due to the non-convexity of~\eqref{eq:P3 trajectory_obj}, we adopt the first-order Taylor expansion to obtain the globally lower bound as follows:
\begin{equation}
\label{eq:Rc SCA}
  \begin{aligned}
  &R^{c,(k)}_{l}\geq\log_2\left(1+\frac{A^{c,(k-1)}_{l}}{B^{c,(k-1)}_{l}+\gamma^{(k)}}\right)-\frac{1}{\ln{2}}\\&\times\frac{\sigma_c^2A^{c,(k-1)}_{l}(\|\mathbf{q}_{c,l}-\mathbf{q}_b\|^2-\|\mathbf{q}_{c,l}^{(k)}-\mathbf{q}_b\|^2)}{(B^{c,(k-1)}_{l,}+\gamma^{(k)})^2+A^{c,(k-1)}_{l}(B^{c,(k-1)}_{l}+\gamma^{(k)})}\\&\triangleq \bar{R}^{c,(k)}_{l},
  \end{aligned}
\end{equation}
where  $A^{c,(k-1)}_{l}=\mathrm{tr}(\mathbf{\tilde{H}}^{bc,(k-1)}_{l}\mathbf{W}_{c,l})$,  $B^{c,(k-1)}_{l}=\mathrm{tr}(\mathbf{\tilde{H}}^{bc,(k-1)}_{l}\mathbf{W}_{e,l})$, $\gamma^{(k)}=\|\mathbf{q}_{c,l}^{(k)}-\mathbf{q}_b\|^2\sigma_c^2$, and $\mathbf{q}_{c,l}^{(k)}$ denotes the given local point at
the $k$-th iteration. Similarly, the non-convex constraint~\eqref{eq:collision} can be relaxed to the convex form by applying the first-order Taylor expansion as follows:
\begin{align}
\label{eq:collision-SCA}
d_\mathrm{min}^{2}&\leq\|\mathbf{q}_{c,l}^{(k)}-\hat{\mathbf{q}}_{e,l|l-1}\|^2\notag\\&+2(\mathbf{q}_{c,l}^{(k)}-\hat{\mathbf{q}}_{e,l|l-1})^T(\mathbf{q}_{c,l}-\mathbf{q}_{c,l}^{(k)}).
\end{align}
Subsequently, in the $k$-th iteration of the SCA, problem~\eqref{eq:P2 trajectory} can be relaxed to the following convex form:
 \begin{subequations}
 \label{eq:trajectory finally}
\begin{equation}
\begin{aligned}
\label{eq:trajectory finally_obj}
    \max_{\mathbf{q}_{c,l}} \bar{R}^{c,(k)}_{l}
\end{aligned}
\end{equation}
\begin{equation}
\label{eq:else 6}
  {\rm{s.t.}} \ \  \eqref{eq:v}, \eqref{eq:field},  \eqref{eq:collision-SCA}.
\end{equation}
 \end{subequations}
Problem~\eqref{eq:trajectory finally} can be readily solved by standard convex solvers, such as CVX~\cite{cvx}. For clarity, the algorithm developed
for solving the C-UAV trajectory optimization problem is outlined in $\textbf{Algorithm~\ref{al_3}}$. 
\begin{algorithm}[tp]
    \renewcommand{\algorithmicrequire}{\textbf{Input:}}
    \renewcommand{\algorithmicensure}{\textbf{Output:}}
  \caption{Proposed Algorithm for Optimizing $\mathbf{q}_{c,l}$}
  \begin{algorithmic}[1]
  \label{al_3}
        \STATE  Initialize feasible point $\mathbf{q}_{c,l}^{(0)}$.  
        \STATE  Set iteration index $k=0$.
      \REPEAT
          \STATE Solve problem $\eqref{eq:trajectory finally}$ to obtain $\mathbf{q}^{(k)}_{c,l}$;
      \STATE Update $\mathbf{q}_{c,l}=\mathbf{q}^{(k)}_{c,l}$ ;
      \STATE $k=k+1$;
      \UNTIL the fractional increment  of $\eqref{eq:trajectory finally}$ is less than the threshold $\mu_2$.
  \end{algorithmic}
\end{algorithm}
\vspace{-0.3cm}
\subsection{Overall Algorithm}
With the solutions obtained for problems~\eqref{eq:beamforming-subproblem} and~\eqref{eq:P1_trajectory}, we now finalize the overall algorithm design for solving the original problem~\eqref{eq:optimization_problem}, which is given in details in \textbf{Algorithm~4}. Specifically, $\textbf{Algorithm~\ref{al_2}}$ and $\textbf{Algorithm~\ref{al_3}}$ are iteratively executed until the increment of the secrecy rate is below the predefined convergence threshold $\mu_4$. Since $\hat{R}_{l}^s$ is non-decreasing during the process in alternatively optimizing $\mathbf{W}_{c,l}$, $\mathbf{W}_{e,l}$, and $\mathbf{q}_{c,l}$, \textbf{Algorithm~\ref{al_4}} is guaranteed to converge to a solution of problem~\eqref{eq:optimization_problem} that is at least locally optimal. 
\begin{table*}[!t]
\caption{Simulation parameters}
\centering
\label{table:simulation-setup}
\begin{tabular}{|m{1.2cm}<{\centering}|m{3.9cm}<{\centering}|m{2.0cm}<{\centering}||m{1.4cm}<{\centering}|m{3.9cm}<{\centering}|m{1.8cm}<{\centering}|}
\hline
Parameter & Description & Value & Parameter & Description & Value\\
\hline
$B$ & Signal bandwidth & $10$~kHz & $\Gamma$ & Maximum E-UAV tracking MSE &$0.2$ \\ \hline
$N$ & Number of symbols in one CPI  & $500$ &$\mathbf{q}_{\mathrm{min}}$ &Flight area lower bound&$\left[-30,-30,25\right]^T$ \\ \hline
$L$ & Total number of CPIs  & $80$ & $\mathbf{q}_{\mathrm{max}}$ & Flight area upper bound &$\left[30,30,40\right]^T$ \\ \hline
$\Delta t$ & CPI duration  & $0.05$~s & $V_{\mathrm{max}}$& Maximum velocity of C-UAV &$5$~m/s\\ \hline
$\beta_0$ & Channel power gain at $1$ meter  & $-30$~dB  & $d_{\mathrm{min}}$ & Minimum safe distance&$7$~m  \\ \hline
$P_{\text{max}}$ & GBS maximum transmit power & $30$~dBm &$\sigma_c^2$,$\sigma_e^2$ & C-UAV and E-UAV noise power  & $-80$~dBm\\ \hline

$\sigma_{\mathrm{RCS}}$ &Radar cross-section of the E-UAV  & $0.03$~$\mathrm{m}^2$ &$\sigma_b^2$ & GBS noise power  & $-50$~dBm  \\ \hline
$c_r$,$c_{\theta}$,$c_{\phi}$ & Scaling factors for location measurement errors & $1$,$10^{-2}$,$10^{-2}$ &$c_{v_r},c_{v_\theta},c_{v_{\phi}}$& Scaling factors for velocity measurement errors & $1$,$1$,$1$ \\ \hline
\end{tabular}
\end{table*}

We now analyze the computational complexity of \textbf{Algorithm 4}. For solving $\mathbf{W}_{c,l}$ and $\mathbf{W}_{e,l}$, the main computational complexity of \textbf{Algorithm~2} relies on applying CVX to solve the SDP problem. If the interior point method is employed, the complexity is given by $\mathcal{O}(2M^{3.5})$. Thus the total complexity is $I_{\mathrm{in}}I_{\mathrm{out}}\left(2M^{3.5}\right)$, where  $I_{\mathrm{in}}$ and $I_{\mathrm{out}}$ represent the required number of inner and outer iterations for convergence, respectively. Similarly, for solving $\mathbf{q}_{c,l}$, \textbf{Algorithm~\ref{al_3}} exhibits the  complexity of $\mathcal{O}(I3^{3.5})$, where $I$ denotes the number of SCA iterations.  Therefore, the overall complexity of $\textbf{Algorithm~\ref{al_4}}$ can be expressed as $\mathcal{O}( I_{\mathrm{AO}}(I_{\mathrm{in}}I_{\mathrm{out}}\left(2M^{3.5}\right)+I3^{3.5}))$, where $I_{\mathrm{AO}}$ denotes the number of iterations required for the convergence of the AO algorithm.

\begin{algorithm}[tp]
    \renewcommand{\algorithmicrequire}{\textbf{Input:}}
    \renewcommand{\algorithmicensure}{\textbf{Output:}}
  \caption{Proposed AO-based algorithm for solving problem $\eqref{eq:optimization_problem}$}
  \begin{algorithmic}[1]
     \label{al_4}
      \STATE  Initialize  $\mathbf{q}_{c,l}^{(0)}$, $\mathbf{W}_{c,l}^{(0)}$ , $\mathbf{W}_{e,l}^{(0)}$.
      \STATE Set iteration index $q = 0$.
      
      \REPEAT
         \STATE Update $\mathbf{W}_{c,l}=\mathbf{W}_{c,l}^{(q)}$ and $\mathbf{W}_{e,l}=\mathbf{W}_{e,l}^{(q)}$ via \textbf{Algorithm 2} with given $\mathbf{q}_{c,l}$;
         \STATE Update $\mathbf{q}_{c,l}=\mathbf{q}_{c,l}^{(q)}$ via \textbf{Algorithm 3} with given $\mathbf{W}_{c,l}$ and $\mathbf{W}_{e,l}$;    

      \STATE $q =q+1$;
      \UNTIL the increment of $\hat{R}_{l}^s$ is below the threshold $\mu_{4}$. 
  \end{algorithmic}
\end{algorithm}

% \textcolor{blue}{$Remark~1:$ This paper focuses on the near-field scenario to fully explore its unique advantages for sensing and security. A core contribution is the 3D velocity sensing scheme (Section III), which exploits the spatial variations of Doppler shifts across the array phenomenon exclusive to the near-field spherical wave model. This enables highly accurate real-time trajectory tracking. However, our framework is inherently adaptable to mixed-field scenarios. If the E-UAV moves into the far-field, the echo signal model degrades to a planar wave front. Consequently, the sensing method would automatically revert to a conventional far-field approach, capable of estimating only the Direction-of-Arrival and radial velocity, losing the 3D velocity information. This degradation in sensing capability is naturally captured by an increased estimation error covariance, $\mathbf{C}_l$. The optimization problem (Section IV) handles this change robustly through the sensing accuracy constraint (34f), which may compel the GBS to allocate more power for sensing to compensate for the reduced accuracy.

\section{Simulation Results}
In this section, numerical results are provided to validate the feasibility and effectiveness of the proposed near-field 3D sensing-aided secure UAV communication framework. We assume that the GBS is equipped with $M = 16 \times 16$ antennas, the system operates at the frequency of $f_c=1.5~\mathrm{GHz}$, and the spacing between the antenna elements is set as half of the operating wavelength. As a result, the corresponding Rayleigh distance is $52$~$\mathrm{m}$. Referring to the $x$-$o$-$y$ plane in Fig.~1, we assume that the E-UAV performs the circular motion centered at $(4.6, 8.0, 28.5)$~meters with the radius of $3.2$~m. Unless otherwise specified, the adopted system parameters are presented in Table~\ref{table:simulation-setup}.

\vspace{-0.3cm}
\subsection{Convergence of Algorithm 4}
\begin{figure}
    \centering
\includegraphics[width=1.0\linewidth]{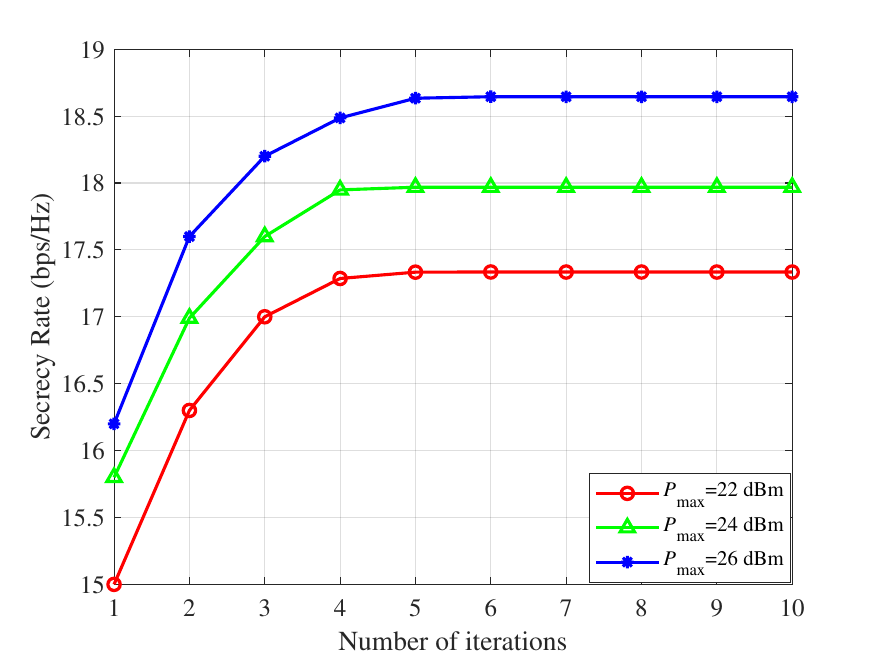}
    \caption{Convergence behavior of the proposed algorithm at one CPI.}
    \label{fig:shoulian}
\end{figure}
In Fig.~\ref{fig:shoulian}, we investigate the convergence behavior of the proposed AO algorithm for the joint GBS beamforming and C-UAV trajectory design under different GBS maximum transmit power $P_{\text{max}}$. The demonstrated results were acquired by running \textbf{Algorithm~\ref{al_4}} in the $18$-th CPI. Note that the secrecy rate is obtained based on~\eqref{eq:secure rate C-UAV}, where $\mathbf{q}^c_{l},\mathbf{w}^{c}_{l},\mathbf{w}^{e}_{l}$ are the output of \textbf{Algorithm 4} given the estimated E-UAV trajectory, while $\mathbf{h}_l^{be}$ in~\eqref{eq:secure rate C-UAV} is calculated based on the E-UAV's ground-truth trajectory.
It is shown that the secrecy rate increases rapidly at the beginning of the iterations and converges within around $5$ iterations. This fast convergence speed implies the possible application of the proposed AO algorithm in practice. It is also observed that the secrecy rate increases with higher GBS maximum transmit powe, which is expected given the higher power gain for the information signal transmitted to the C-UAV and the enhanced AN signal strength at the E-UAV.  
\subsection{E-UAV Trajectory Tracking and C-UAV Trajectory Design Performance}
In Fig.~\ref{E-UAV and C-UAV trajectory}, we depict the estimated E-UAV trajectory obtained with the trajectory tracking scheme designed in Section III as well as the C-UAV trajectory obtained with \textbf{Algorithm~\ref{al_4}} given different initial locations. Specifically, we set three different C-UAV's initial locations, given by $[10,2,30]^T$, $[14,6,30]^T$, $[12,16,30]^T$, and label the corresponding trajectory as ``C-UAV trajectory 1", ``C-UAV trajectory 2", and ``C-UAV trajectory 3", respectively. The E-UAV's ground-truth trajectory is also plotted for comparison. 
One can first observe that estimated trajectory perfectly aligns with the ground-truth trajectory of the E-UAV, which underscores the high accuracy for tracking the E-UAV trajectory with the proposed near-field 3D sensing method. Observing the C-UAV's trajectories, one can find that the C-UAV's flight tends to approach the GBS, which is expected as the secrecy rate can be improved when the path loss between the GBS and C-UAV is reduced. Moreover, it is also seen that the C-UAV keeps adjusting its flight during the given time period rather than adopting the straight-line trajectory. The is because, due to the presence of the E-UAV, the C-UAV trajectory should be designed to guarantee the minimum safe distance, as well as to reduce the channel correlations with the E-UAV so as to improve the secrecy rate. From the 3D trajectory plot given in Fig.~6(b), we also see that the C-UAV initially descends rapidly along the z-axis until reaches to the predetermined minimum flight altitude, after which the height is not changed anymore. This is quite intuitive as the C-UAV tends to be get close to the GBS to improve the achievable rate. 

\begin{figure}
	\centering
	\begin{subfigure}{1\linewidth}
		\centering
		\includegraphics[width=1.0\linewidth]{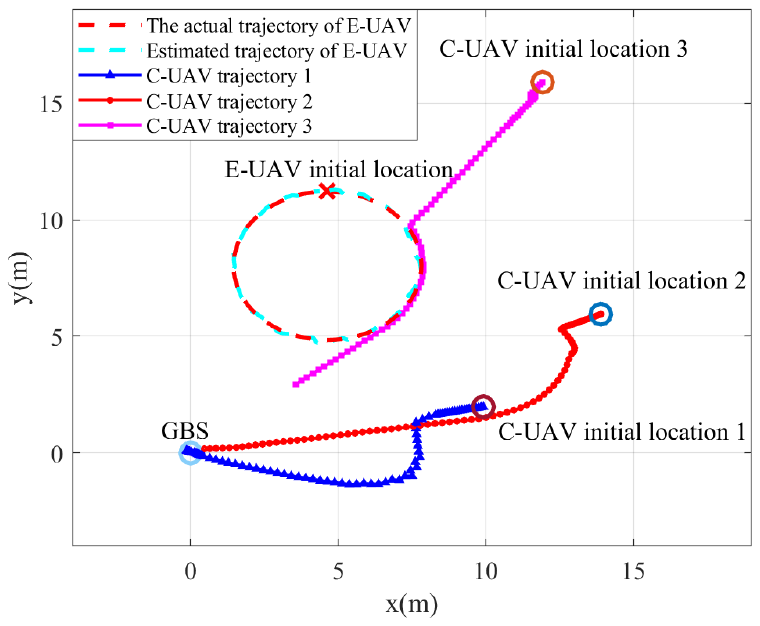}
		\caption{2D-space demonstration.}
		\label{2D}%文中引用该图片代号
	\end{subfigure}
	\centering
	\begin{subfigure}{1\linewidth}
		\centering
		\includegraphics[width=1.1\linewidth]{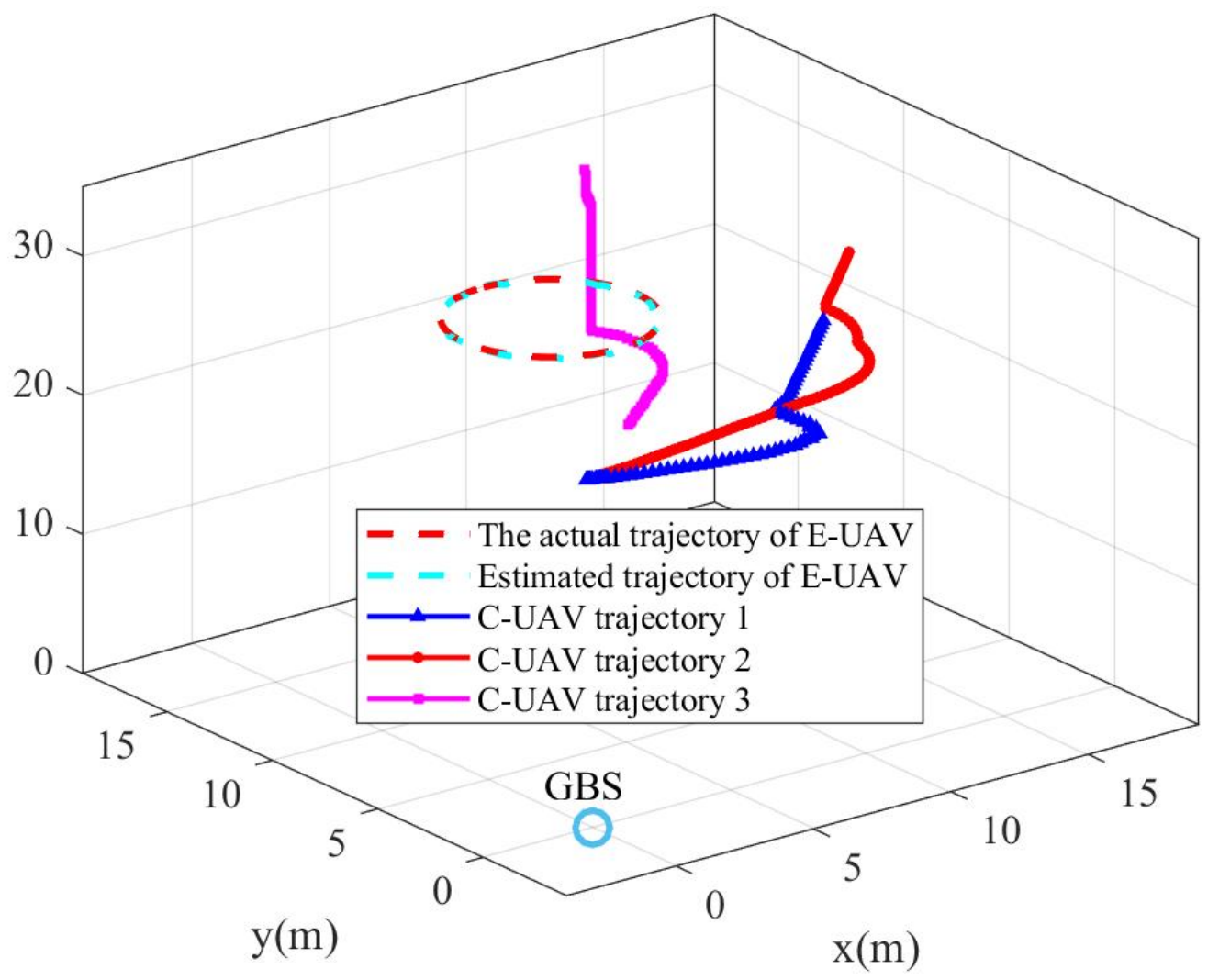}
		\caption{3D-space demonstration.}
		\label{3D}%文中引用该图片代号
	\end{subfigure}
    \caption{Demonstration of the E-UAV's estimated trajectory and the C-UAV's designed trajectory.}
	\label{E-UAV and C-UAV trajectory}
\end{figure}

\begin{figure}
    \centering
    \includegraphics[width=1.0\linewidth]{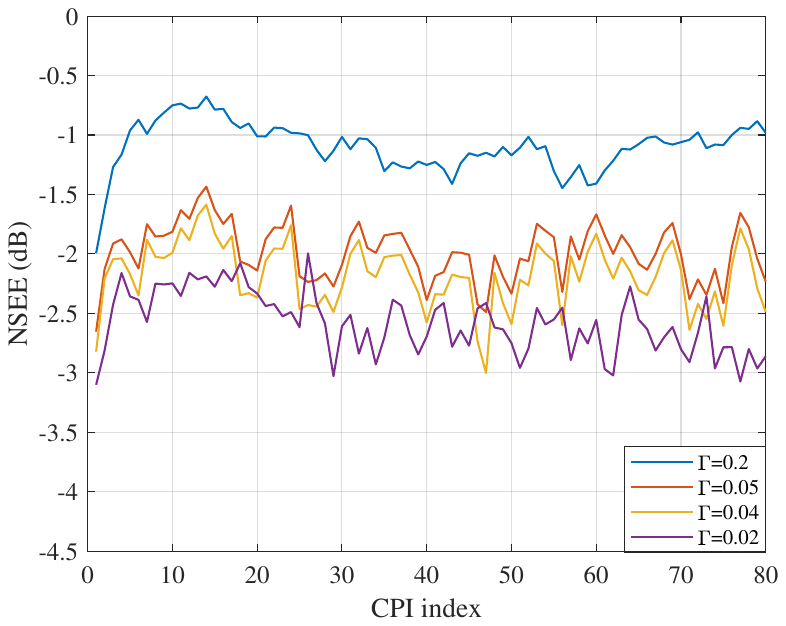}
    \caption{Instantaneous normalized E-UAV's state estimation error versus the maximum tracking MSE.}
    \label{fig:wucha}
\end{figure}
In Fig.~\ref{fig:wucha}, we plot the instantaneous normalized E-UAV's state estimation error (NSEE) during the whole flight duration of $80$ CPIs. The NSEE is defined as $\mathrm{log_{10}(\frac{|\mathbf{s}^{e}_{l}-\hat{\mathbf{s}}^{e}_{l}|^2}{\rho})}$, with the normalization factor $\rho$ set to $10$. 
It is evident to see that the NSEE increases with the increment of the maximum E-UAV tracking MSE $\Gamma$. Specifically, define the average NSEE over $L$ CPIs as $\frac{1}{L}\sum_{l=1}^{L}\mathrm{log_{10}(\frac{|\mathbf{s}^{e}_{l}-\hat{\mathbf{s}}^{e}_{l}|^2}{\rho})}$. We obtain the average NSEE over the $80$ CPIs as $-1.09$~dB, $-1.98$~dB, $-2.18$~dB, and $-2.58$~dB for $\Gamma$ given by $0.2$, $0.05$, $0.04$, and $0.02$, respectively. This is expected, as the instantaneous E-UAV's state estimation error gets smaller when the estimation accuracy requirement becomes more strict. We also observe that the instantaneous NSEE provided in Fig.~\ref{fig:wucha} does not monotonically decrease over time, which is due to the E-UAV's varying locations during the whole time period. 

\subsection{Secure UAV Communications Performance}
We choose the ``C-UAV trajectory 1" in Fig.~6 for demonstrating the secure UAV communication performance in this subsection. In Fig.~8, we depict the instantaneous secrecy rate with different GBS maximum transmit power $P_{\text{max}}$. One can observe that the secrecy rate increases rapidly at the beginning of the flight, which is due to the fact that the C-UAV quickly descends and gets closer to the GBS. Subsequently, the C-UAV adjusts its flight direction to maintain a safe distance from the E-UAV, resulting in a slower increment of the secrecy rate. Once the C-UAV reaches to the position that is directly above the GBS, the secrecy rate remains approximately constant. 

\begin{figure}
    \centering
    \includegraphics[width=1\linewidth]{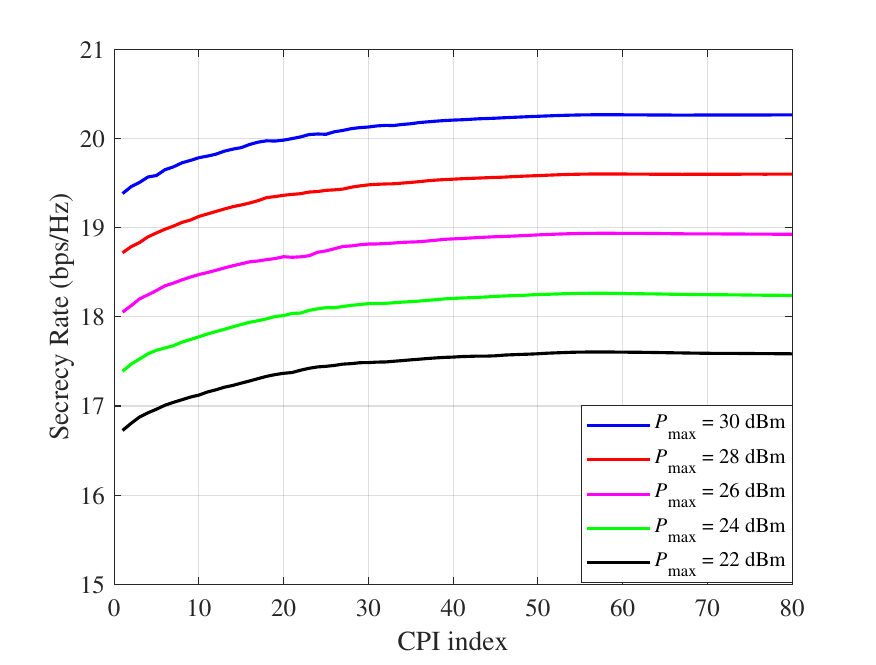}
    \caption{Secrecy rate versus GBS maximum transmit power.}
    \label{fig:P}
\end{figure}

In Fig.~\ref{fig:s_mse}, we present the instantaneous secrecy rate with different maximum E-UAV tracking MSE $\Gamma$. It is interesting to find that, the secrecy rate does not change much when  $\Gamma$ is relatively large. However, a significant decrement of the secrecy rate is observed when $\Gamma$ drops below $0.04$, which can be explained as follows. When $\Gamma$ gets smaller, the GBS needs to allocate more power to the AN signals, which results in lower power allocation to the information signal given that the GBS total transmit power is limited. This performance degradation gets more severe when $l$ is larger than $50$. This is because, after $50$ CPIs, the E-UAV moves farther from the GBS while the C-UAV's location keeps almost static as it has reached to the location above the GBS. Thus, the GBS needs to allocate more power to the AN signals to guarantee the sensing accuracy, which leads to the decrement of secrecy rate as the power allocated to the C-UAV gets lower. These observations highlight the trade-off between the E-UAV tracking accuracy and the secure communication performance. In other words, to achieve high secrecy rate, the sensing accuracy threshold should be given a proper value. 
\begin{figure}
    \centering
    \includegraphics[width=1.0\linewidth]{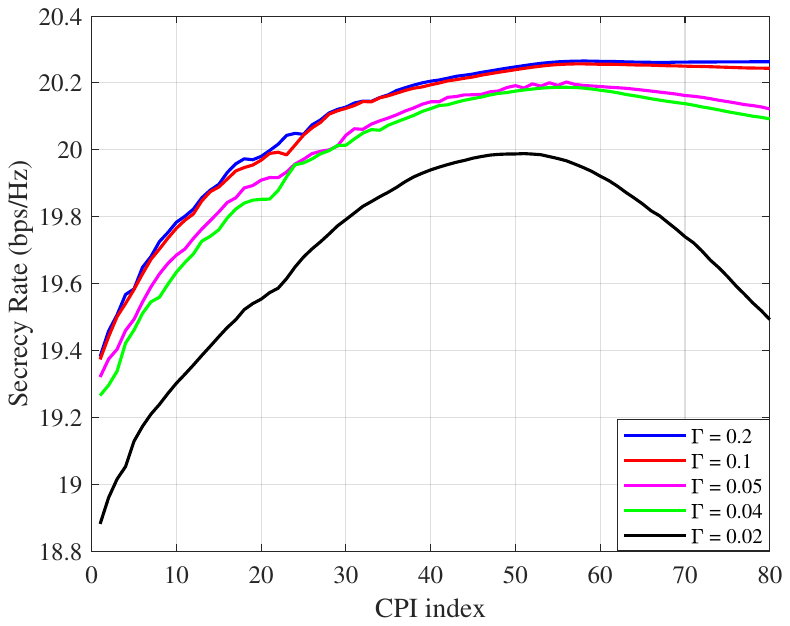}
    \caption{Secrecy rate versus maximum E-UAV tracking MSE.}
    \label{fig:s_mse}
\end{figure}

\begin{figure}
    \centering
    \includegraphics[width=1.0\linewidth]{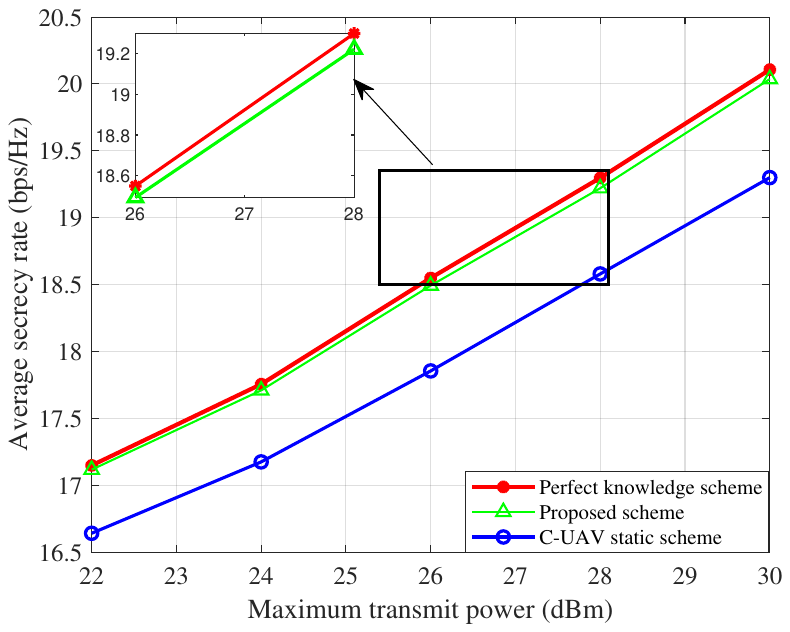}
    \caption{Average secrecy rate versus maximum transmit power.}
    \label{fig:duibi-p}
\end{figure}

\begin{figure}
    \centering
    \includegraphics[width=1.0\linewidth]{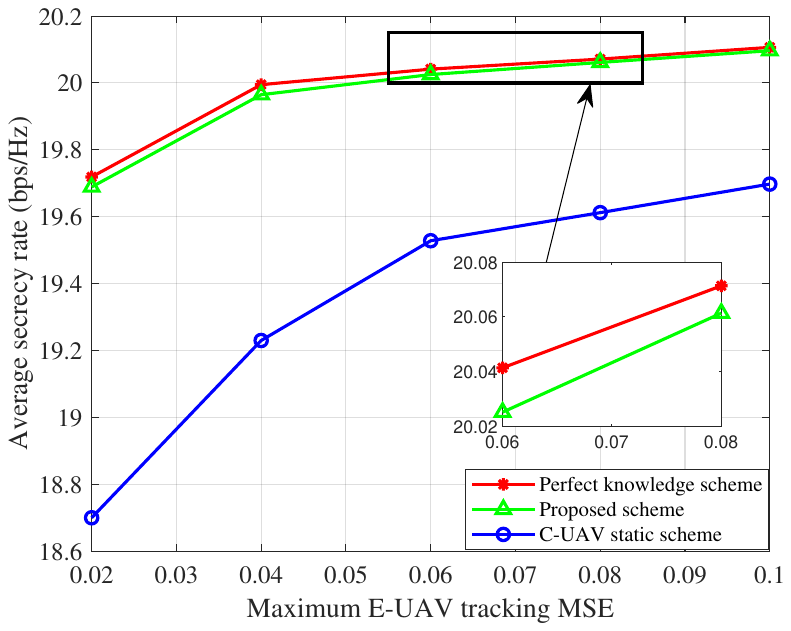}
    \caption{Average secrecy rate versus maximum E-UAV tracking MSE.}
    \label{fig:duibi-s}
\end{figure}

\subsection{Comparison with Baseline Schemes}
To further evaluate the effectiveness of the proposed near-field sensing-aided secure UAV communication framework, we compare with the following two baseline schemes:
\begin{itemize}
\item  \textbf{Perfect knowledge scheme} (denoted as ``\textbf{PKS}"): In this case, the E-UAV trajectory is assumed to be perfectly known at the GBS. Therefore, the resulting joint GBS beamforming and C-UAV trajectory design problem can be solved by applying \textbf{Algorithm~\ref{al_4}} with the precise E-UAV's real-time locations. 
\item \textbf{C-UAV static scheme} (denoted as ``\textbf{CSS}"): In this case, the location of the C-UAV is assume to be fixed at the point $(12.0,3.0,10.0)$~meters, while the GBS beamforming is designed by applying \textbf{Algorithm~\ref{al_2}}.
\end{itemize}

In Fig.~\ref{fig:duibi-p}, we compare the average secrecy rate over $80$ CPIs achieved by different schemes versus the GBS maximum transmit power. As can be observed, the performance of the proposed scheme nearly approaches to that of the ``PKS", which confirms the effectiveness of the proposed E-UAV trajectory tracking scheme. Moreover, the proposed scheme is shown to outperform the ``CSS", which implies that the secure communication performance can be improved by properly designing the C-UAV trajectory with \textbf{Algorithm~\ref{al_4}}.

In Fig.~\ref{fig:duibi-s}, we further provide the performance comparison between the proposed scheme and the baselines versus the maximum E-UAV tracking MSE $\Gamma$. 
It is shown that the performance gap between the proposed scheme and the ``PKS" is negligible. Moreover, we can observe that the performance gap between the proposed scheme and the ``CSS" is much significant when $\Gamma$ is small. This is because, the GBS needs to allocate more power to the AN signals when the sensing accuracy requirement is strict. As the distance between the GBS and the C-UAV is large for ``CSS", the secrecy rate degradation is more pronounced when less power is allocated to the information signal. 
\vspace{-0.3cm}
\section{Conclusion}
In this paper, a novel near-field 3D sensing-aided secure UAV communication framework with high time efficiency was proposed, where the GBS sent the AN signals for both the sensing and jamming purpose to the E-UAV. 
To obtain the unknown E-UAV's locations in real time, an E-UAV trajectory tracking scheme was first introduced. Specifically, the 3D velocities were estimated with the echo signals by exploiting the variant Doppler shift observations over the spatial domain in the near field. An EKF-based data fusion method was further designed to timely correct the E-UAV localization errors. Based on the E-UAV's predicted location with the measured 3D velocities, a joint GBS beamforming and C-UAV trajectory design scheme was proposed for maximizing the instantaneous secrecy rate, while guaranteeing the sensing accuracy. An AO algorithm was developed for solving the resulting non-convex problem, where the subproblems were iteratively solved with the SCA method.  Extensive simulation results were provided to verify the superiority of the proposed scheme. It was shown that the proposed E-UAV trajectory tracking scheme can achieve accurate localization of the E-UAV in real time. Moreover, the secrecy rate achieved by the proposed scheme was demonstrated to closely approaches to that obtained by the case where the E-UAV trajectory was perfectly known.

%Note that the self-interference caused by simultaneous transmission and reception during sensing is assumed to be cancelled by employing proper self-interference cancellation methods.
\vspace{-0.4cm}
\bibliographystyle{IEEEtran} 
\bibliography{mybib}

\begin{thebibliography}{10}
\providecommand{\url}[1]{#1}
\csname url@samestyle\endcsname
\providecommand{\newblock}{\relax}
\providecommand{\bibinfo}[2]{#2}
\providecommand{\BIBentrySTDinterwordspacing}{\spaceskip=0pt\relax}
\providecommand{\BIBentryALTinterwordstretchfactor}{4}
\providecommand{\BIBentryALTinterwordspacing}{\spaceskip=\fontdimen2\font plus
\BIBentryALTinterwordstretchfactor\fontdimen3\font minus \fontdimen4\font\relax}
\providecommand{\BIBforeignlanguage}[2]{{%
\expandafter\ifx\csname l@#1\endcsname\relax
\typeout{** WARNING: IEEEtran.bst: No hyphenation pattern has been}%
\typeout{** loaded for the language `#1'. Using the pattern for}%
\typeout{** the default language instead.}%
\else
\language=\csname l@#1\endcsname
\fi
#2}}
\providecommand{\BIBdecl}{\relax}
\BIBdecl

\bibitem{UAV1}
Y.~Zeng, Q.~Wu, and R.~Zhang, ``Accessing from the sky: A tutorial on {UAV} communications for {5G} and beyond,'' \emph{Proc.{IEEE}}, vol. 107, no.~12, pp. 2327--2375, Dec. 2019.

\bibitem{UAV2}
R.~Zhang, H.~Du, D.~Niyato, J.~Kang, Z.~Xiong, A.~Jamalipour, P.~Zhang, and D.~I. Kim, ``Generative {AI} for space-air-ground integrated networks,'' \emph{{IEEE} Wireless Commun.}, vol.~31, no.~6, pp. 10--20, Dec. 2024.

\bibitem{UAVLOS1}
J.~Zhao, Q.~Xu, X.~Mu, Y.~Liu, and Y.~Zhu, ``Aerial active {STAR-RIS}-aided {IoT NOMA} networks,'' \emph{{IEEE} Internet Things J.}, vol.~12, no.~8, pp. 9525--9538, Apr. 2025.

\bibitem{site}
R.~Zhang, H.~Du, Y.~Liu, D.~Niyato, J.~Kang, Z.~Xiong, A.~Jamalipour, and D.~In~Kim, ``Generative {AI} agents with large language model for satellite networks via a mixture of experts transmission,'' \emph{{IEEE} J. Sel. Areas Commun.}, vol.~42, no.~12, pp. 3581--3596, Dec. 2024.

\bibitem{UAVLOS2}
B.~Duo, Q.~Wu, X.~Yuan, and R.~Zhang, ``Anti-jamming {3D} trajectory design for {UAV}-enabled wireless sensor networks under probabilistic {LoS} channel,'' \emph{{IEEE} Trans. Veh. Technol.}, vol.~69, no.~12, pp. 16\,288--16\,293, Dec. 2020.

\bibitem{MIMOUAV1}
J.~Zhao, J.~Su, K.~Cai, Y.~Zhu, Y.~Liu, and N.~Al-Dhahir, ``Interference-robust broadband rapidly-varying {MIMO} communications: A knowledge-data dual driven framework,'' \emph{{IEEE} Trans. Wireless Commun.}, pp. 1--17, Mar. 2025, early access, doi: {\color{blue} {10.1109/TWC.2025.3548618}}.

\bibitem{MIMOSECURE}
M.~Jiang, Y.~Li, Q.~Zhang, Q.~Li, and J.~Qin, ``Secure beamforming in downlink {MIMO} nonorthogonal multiple access networks,'' \emph{{IEEE} Signal Process. Lett.}, vol.~24, no.~12, pp. 1852--1856, Dec. 2017.

\bibitem{pilot1}
W.~Ma, C.~Qi, and G.~Y. Li, ``High-resolution channel estimation for frequency-selective mmwave massive {MIMO} systems,'' \emph{{IEEE} Trans. Commun. Technol.}, vol.~19, no.~5, pp. 3517--3529, May 2020.

\bibitem{9737357}
F.~Liu, Y.~Cui, C.~Masouros, J.~Xu, T.~X. Han, Y.~C. Eldar, and S.~Buzzi, ``Integrated sensing and communications: Toward dual-functional wireless networks for {6G} and beyond,'' \emph{{IEEE} J. Sel. Areas Commun.}, vol.~40, no.~6, pp. 1728--1767, June 2022.

\bibitem{9851407}
T.~Mao, J.~Chen, Q.~Wang, C.~Han, Z.~Wang, and G.~K. Karagiannidis, ``Waveform design for joint sensing and communications in {Millimeter-Wave} and low {Terahertz} bands,'' \emph{{IEEE} Trans. Commun.}, vol.~70, no.~10, pp. 7023--7039, Oct. 2022.

\bibitem{ISAC1}
Z.~Wei, H.~Qu, Y.~Wang, X.~Yuan, H.~Wu, Y.~Du, K.~Han, N.~Zhang, and Z.~Feng, ``Integrated sensing and communication signals toward {5G-A} and {6G}: {A} survey,'' \emph{{IEEE} Internet Things J.}, vol.~10, no.~13, pp. 11\,068--11\,092, Jul. 2023.

\bibitem{ISAC2}
J.~Wang, N.~Varshney, C.~Gentile, S.~Blandino, J.~Chuang, and N.~Golmie, ``Integrated sensing and communication: Enabling techniques, applications, tools and data sets, standardization, and future directions,'' \emph{{IEEE} Internet Things J.}, vol.~9, no.~23, pp. 23\,416--23\,440, Dec. 2022.

\bibitem{ISAC3}
S.~Lu, F.~Liu, Y.~Li, K.~Zhang, H.~Huang, J.~Zou, X.~Li, Y.~Dong, F.~Dong, J.~Zhu, Y.~Xiong, W.~Yuan, Y.~Cui, and L.~Hanzo, ``Integrated sensing and communications: {Recent} advances and ten open challenges,'' \emph{{IEEE} Internet Things J.}, vol.~11, no.~11, pp. 19\,094--19\,120, Jun. 2024.

\bibitem{NFCMIMO1}
H.~Lu, Y.~Zeng, C.~You, Y.~Han, J.~Zhang, Z.~Wang, Z.~Dong, S.~Jin, C.-X. Wang, T.~Jiang, X.~You, and R.~Zhang, ``A tutorial on near-field {XL-MIMO} communications toward {6G},'' \emph{{IEEE} Commun. Surveys Tuts.}, vol.~26, no.~4, pp. 2213--2257, early access, Apr. 12, 2024.

\bibitem{NFCMIMO2}
X.~Mu, J.~Xu, Y.~Liu, and L.~Hanzo, ``Reconfigurable intelligent surface-aided near-field communications for {6G}: Opportunities and challenges,'' \emph{{IEEE} Veh. Technol. Mag.}, vol.~19, no.~1, pp. 65--74, Mar. 2024.

\bibitem{UAVSECURE2}
A.~V. Savkin, H.~Huang, and W.~Ni, ``Securing {UAV} communication in the presence of stationary or mobile eavesdroppers via online {3D} trajectory planning,'' \emph{{IEEE} Wireless Commun. Lett.}, vol.~9, no.~8, pp. 1211--1215, Aug. 2020.

\bibitem{UAVSECURE1}
M.~Cui, G.~Zhang, Q.~Wu, and D.~W.~K. Ng, ``Robust trajectory and transmit power design for secure {UAV} communications,'' \emph{{IEEE} Trans. Veh. Technol.}, vol.~67, no.~9, pp. 9042--9046, Sep. 2018.

\bibitem{UAVSECURE3}
Z.~Na, C.~Ji, B.~Lin, and N.~Zhang, ``Joint optimization of trajectory and resource allocation in secure {UAV} relaying communications for {Internet} of things,'' \emph{{IEEE} Internet Things J.}, vol.~9, no.~17, pp. 16\,284--16\,296, Sep. 2022.

\bibitem{UAVJAMMING2}
Y.~Zhou, P.~L. Yeoh, C.~Pan, K.~Wang, Z.~Ma, B.~Vucetic, and Y.~Li, ``Caching and {UAV} friendly jamming for secure communications with active eavesdropping attacks,'' \emph{{IEEE} Trans. Veh. Technol.}, vol.~71, no.~10, pp. 11\,251--11\,256, Oct. 2022.

\bibitem{multiuav}
R.~Li, Z.~Wei, L.~Yang, D.~W.~K. Ng, J.~Yuan, and J.~An, ``Resource allocation for secure multi-{UAV} communication systems with multi-eavesdropper,'' \emph{{IEEE} Trans. Commun.}, vol.~68, no.~7, pp. 4490--4506, Jul. 2020.

\bibitem{9416239}
S.~Li, B.~Duo, M.~D. Renzo, M.~Tao, and X.~Yuan, ``Robust secure {UAV} communications with the aid of reconfigurable intelligent surfaces,'' \emph{{IEEE} Wireless Commun. Lett.}, vol.~20, no.~10, pp. 6402--6417, Oct. 2021.

\bibitem{10288199}
R.~Ye, Y.~Peng, F.~Al-Hazemi, and R.~Boutaba, ``A robust cooperative jamming scheme for secure {UAV} communication via intelligent reflecting surface,'' \emph{{IEEE} Trans. Commun.}, vol.~72, no.~2, pp. 1005--1019, Feb. 2024.

\bibitem{ISACFAR1}
J.~Chu, R.~Liu, M.~Li, Y.~Liu, and Q.~Liu, ``Joint secure transmit beamforming designs for integrated sensing and communication systems,'' \emph{{IEEE} Trans. Veh. Technol.}, vol.~72, no.~4, pp. 4778--4791, Apr. 2023.

\bibitem{ISACFAR2}
Z.~Ren, L.~Qiu, J.~Xu, and D.~W.~K. Ng, ``Robust transmit beamforming for secure integrated sensing and communication,'' \emph{{IEEE} Trans. Commun.}, vol.~71, no.~9, pp. 5549--5564, Sep. 2023.

\bibitem{ISACFAR3}
N.~Su, F.~Liu, and C.~Masouros, ``Sensing-assisted eavesdropper estimation: An {ISAC} breakthrough in physical layer security,'' \emph{{IEEE} Trans. Wireless Commun.}, vol.~23, no.~4, pp. 3162--3174, Apr. 2024.

\bibitem{ISACli2}
X.~Meng, F.~Liu, C.~Masouros, W.~Yuan, Q.~Zhang, and Z.~Feng, ``Vehicular connectivity on complex trajectories: Roadway-geometry aware {ISAC} beam-tracking,'' \emph{{IEEE} Trans. Wireless Commun.}, vol.~22, no.~11, pp. 7408--7423, Mar. 2023.

\bibitem{ISACli1}
F.~Xia, Z.~Fei, J.~Huang, X.~Wang, R.~Wang, W.~Yuan, and D.~W.~K. Ng, ``Sensing-enabled predictive beamforming design for {RIS}-assisted {V2I} systems: A deep learning approach,'' \emph{{IEEE} Trans. Wireless Commun.}, vol.~23, no.~6, pp. 5571--5586, Jun. 2024.

\bibitem{weizhiqiang}
Z.~Wei, F.~Liu, C.~Liu, Z.~Yang, D.~W.~K. Ng, and R.~Schober, ``Integrated sensing, navigation, and communication for secure {UAV} networks with a mobile eavesdropper,'' \emph{{IEEE} Trans. Wireless Commun.}, vol.~23, no.~7, pp. 7060--7078, Jul. 2024.

\bibitem{NFC-LOCA2}
X.~Cao, A.~Chen, H.~Yin, and L.~Chen, ``A low-complexity near-field localization method based on electric field model,'' \emph{{IEEE} Commun. Lett.}, vol.~28, no.~2, pp. 288--292, Dec. 2024.

\bibitem{10288339}
K.~Chen, C.~Qi, C.-X. Wang, and G.~Y. Li, ``Beam training and tracking for extremely large-scale {MIMO} communications,'' \emph{{IEEE} Trans. Wireless Commun.}, vol.~23, no.~5, pp. 5048--5062, May 2024.

\bibitem{senseaid}
D.~Galappaththige, S.~Zargari, C.~Tellambura, and G.~Y. Li, ``Near-field {ISAC}: Beamforming for multi-target detection,'' \emph{{IEEE} Wireless Commun. Lett.}, vol.~13, no.~7, pp. 1938--1942, Jul. 2024.

\bibitem{NFC_V}
Z.~Wang, X.~Mu, and Y.~Liu, ``Near-field velocity sensing and predictive beamforming,'' \emph{{IEEE} Trans. Veh. Technol.}, vol.~74, no.~1, pp. 1806--1810, Jan. 2025.

\bibitem{DULL}
A.~Sabharwal, P.~Schniter, D.~Guo, D.~W. Bliss, S.~Rangarajan, and R.~Wichman, ``In-band full-duplex wireless: Challenges and opportunities,'' \emph{{IEEE} J. Sel. Areas Commun.}, vol.~32, no.~9, pp. 1637--1652, Sep. 2014.

\bibitem{zaboyizhi}
M.~I. Skolnik, \emph{Introduction to RADAR systems}.\hskip 1em plus 0.5em minus 0.4em\relax Introduction to RADAR systems, 1990.

\bibitem{10078317}
Y.~Lu and L.~Dai, ``Near-field channel estimation in mixed {LoS/NLoS} environments for extremely large-scale {MIMO} systems,'' \emph{{IEEE} Trans. Commun.}, vol.~71, no.~6, pp. 3694--3707, June 2023.

\bibitem{MF}
S.~K. Sengijpta, ``Fundamentals of statistical signal processing: Estimation theory,'' \emph{Control Engineering Practice}, vol.~37, no.~4, pp. 465--466, 1994.

\bibitem{celiangwucha}
C.~Richmond, ``Mean-squared error and threshold {SNR} prediction of maximum-likelihood signal parameter estimation with estimated colored noise covariances,'' \emph{{IEEE} Trans. Inf. Theory}, vol.~52, no.~5, pp. 2146--2164, May 2006.

\bibitem{snr_c}
F.~Liu, W.~Yuan, C.~Masouros, and J.~Yuan, ``Radar-assisted predictive beamforming for vehicular links: Communication served by sensing,'' \emph{{IEEE} Trans. Wireless Commun.}, vol.~19, no.~11, pp. 7704--7719, Nov. 2020.

\bibitem{EKF}
B.~D.~O. Anderson, J.~B. Moore, and M.~Eslami, \emph{Optimal Filtering}.\hskip 1em plus 0.5em minus 0.4em\relax Courier Corporation, 2012.

\bibitem{Schur}
F.~Z. Zhang, \emph{The Schur Complement and Its Applications}.\hskip 1em plus 0.5em minus 0.4em\relax Springer US, 2005.

\bibitem{penalty}
\BIBentryALTinterwordspacing
A.~Ben-Tal and M.~Zibulevsky, ``Penalty/barrier multiplier methods for convex programming problems,'' \emph{SIAM Journal on Optimization}, vol.~7, no.~2, pp. 347--366, 1997. [Online]. Available: \url{https://doi.org/10.1137/S1052623493259215}
\BIBentrySTDinterwordspacing

\bibitem{boyd2004convex}
S.~Boyd and L.~Vandenberghe, \emph{Convex optimization}.\hskip 1em plus 0.5em minus 0.4em\relax Cambridge, U.K.: Cambridge Univ. Press, 2004.

\bibitem{k+1_k}
S.~Li, B.~Duo, M.~D. Renzo, M.~Tao, and X.~Yuan, ``Robust secure uav communications with the aid of reconfigurable intelligent surfaces,'' \emph{{IEEE} Trans. Wireless Commun.}, vol.~20, no.~10, pp. 6402--6417, Oct. 2021.

\bibitem{cvx}
M.~Grant and S.~Boyd, ``{CVX}: Matlab software for disciplined convex programming, version 2.1,'' \url{https://cvxr.com/cvx}, Mar. 2014.

\end{thebibliography}
\end{document}